\documentclass[12pt,notoc]{JHEP3}

\usepackage{amsmath,amssymb,euscript,array,cite,mathrsfs,amsfonts}
\usepackage{epsfig}

\newcommand{\field}[1]{\mathbb{#1}} 
\newcommand*{\Dsl}[0]{{\rlap{\kern2.25pt /}{D}}}
\newcommand*{\Asl}[0]{{\rlap{\kern2.25pt /}{A}}}
\newcommand*{\dsl}[0]{{\rlap{\kern0.5pt /}{\partial}}}
\newcommand*{\xisl}[0]{{\rlap{\kern0.5pt /}{\xi}}}
\newcommand*{\asl}[0]{{\rlap{\kern0.5pt /}{a}}}
\newcommand*{\bsl}[0]{{\rlap{\kern0.5pt /}{b}}}

\newcommand{\lapprox}{\raisebox{-0.5ex}{$\
\stackrel{\textstyle<}{\textstyle\sim}\ $}}
\newcommand{\gapprox}{\raisebox{-0.5ex}{$\
\stackrel{\textstyle>}{\textstyle\sim}\ $}}

\def\Dslash{\,\,{\raise.15ex\hbox{/}\mkern-12mu D}}

\newcommand{\SP}[1]{\begin{equation}\begin{split} #1
\end{split}\end{equation}}

\def\kN{{\cal N}}

\newcommand{\Tr}{\operatorname{Tr}}
\newcommand{\RE}{\operatorname{Re}}

\def\B0{{\boldsymbol 0}}

\def\Tr{{\rm Tr}}

\def\det{{\rm det}}

\def\Dbarslash{\,\,{\raise.15ex\hbox{/}\mkern-12mu {\bar D}}}
\def\Dslash{\,\,{\raise.15ex\hbox{/}\mkern-12mu D}}
\def\delslash{\,\,{\raise.15ex\hbox{/}\mkern-9mu \partial}}
\def\delbarslash{\,\,{\raise.15ex\hbox{/}\mkern-9mu {\bar\partial}}}

\newcommand{\IM}{\operatorname{Im}}

\def\LAG{\mathscr{L}}

\def\PP{\mathscr{P}}

\newcommand{\EQ}[1]{\begin{equation}\begin{split} #1
\end{split}\end{equation}}

\title{QCD with Chemical Potential in a Small Hyperspherical Box}
\author{Simon Hands, Timothy J. Hollowood and Joyce C. Myers\\ Swansea University,
  Physics Department, Swansea SA2 8PP, UK\\ E-mail:
  \email{s.hands@swansea.ac.uk, t.hollowood@swansea.ac.uk, j.c.myers@swansea.ac.uk}}

\abstract{To leading order in perturbation theory, 
we solve QCD, defined on a small three sphere
in the large $N$ and $N_f$ limit, at finite chemical
  potential and map out the phase diagram in the $(\mu,T)$ plane.
The action of QCD is complex in the presence of
  a non-zero quark chemical potential which results in the sign
  problem for lattice simulations. 
In the large $N$ theory, which at low temperatures becomes
  a conventional unitary matrix model with a complex action, 
we find that the dominant contribution to
  the functional integral comes from complexified gauge field
  configurations. For this reason the eigenvalues of the Polyakov line
  lie off the unit circle on a contour in the complex
  plane. We find at low temperatures that as $\mu$ passes one of the 
quark energy
  levels there is a third-order Gross-Witten transition from a confined 
to a deconfined phase and back again giving rise to a rich phase structure.
We compare a range of physical observables in the large $N$ theory to
those calculated numerically in the theory with $N=3$. In the latter
case there are no genuine phase transitions in a finite volume but nevertheless the
observables are remarkably similar to the large $N$ theory.}


\preprint{}

\begin{document}


\section{Introduction}

The phase diagram of QCD at finite quark number density is of considerable
interest but a first principles derivation of the grand potential has
eluded us for two very good reasons. The first is that the phase
transitions are conjectured to occur at densities where the 
coupling strength of QCD is large and thus the application of
conventional perturbation theory is not
valid. The other reason is that the action of QCD is complex
resulting in the {\it sign problem\/}. This prevents the use of
the usual technique of importance sampling in lattice gauge theory
simulations as it is not possible to formulate a probability
interpretation with a complex Boltzmann factor. However, there has been
progress since several ways to work around the sign
problem have been discovered, as will be discussed shortly. 
We propose a complementary idea which is to formulate
the theory on a manifold with sufficiently small spatial volume such
that perturbation theory is valid, and so that the calculation is
valid at all temperatures and densities. 
Our particular choice of spatial manifold 
is $S^3$, which is motivated by the connection with the AdS/CFT
correspondence. In a nut-shell, when the 
maximally super-symmetric ${\cal N}=4$ gauge theory is defined in a
compact space, an $S^3$, its thermodynamics in the large $N$ limit
and at strong 't~Hooft coupling can be addressed by the dual
gravitational description. What is remarkable is that the phase
structure seems to match on to the weak coupling description that is
addressed in perturbation theory; namely, as the temperature is raised
there is a confinement/deconfinement type transition, a phase
transition that is described in the gravity dual as a Hawking-Page
transition from AdS space to an AdS black hole space 
\cite{Witten:1998zw,Sundborg:1999ue,Aharony:2003sx}. This
kind of analysis has been extended to include an $R$-charge chemical potential
\cite{Yamada:2006rx}. What we take from this is that it is interesting to
investigate the phase structure of any gauge theory in finite volume,
on an $S^3$, at weak coupling which is ensured if the size of the
$S^3$, $R$, is  much smaller than the strong coupling scale
$R\ll\Lambda_{QCD}^{-1}$. In this case a thermodynamic limit is ensured by
taking the large $N$ limit and genuine phase transitions occur. These
transitions have all the characteristics of phase transitions that are
expected in theories with finite $N$ defined on flat space. The goal
of the present paper is to extend this kind of analysis to $SU(N)$
gauge theories with $N_f$ fundamental quarks in the large $N$
Veneziano limit \cite{Veneziano:1976wm}, {\it i.e.\/}~with 
the ratio $\frac{N_f}N$
fixed, with both finite temperature and baryon chemical potential. It will be interesting to relate our weak coupling results to
strong coupling analysis based on the AdS/CFT correspondence. This
will involve adding ``flavour branes'' to the basic set
up to describe the quarks.\footnote{The process of 
adding flavour to the basic AdS/CFT set-up
  has a huge literature. Most of this work addresses the case where the
  boundary theory is defined on ${\boldsymbol R}^3$ and with fixed
  $N_f$, so that the flavours can be introduced in the probe
  approximation; however, the papers \cite{Karch:2006bv,Karch:2009ph}
  consider the case of global $AdS$ with an $S^3$ boundary.}

Our approach should have implications for understanding
one of the outstanding problems in theoretical physics; namely,
the behaviour of cold dense baryonic matter, which in essence
corresponds to an understanding of QCD with non-zero baryon
chemical potential $\mu$.
Its resolution would permit contact between particle and nuclear physics via a
quantitative description of bulk nuclear matter from first principles, and would
set the study of compact stars on a firm theoretical footing, via
the input of the QCD equation of state (energy density $\varepsilon(\mu)$,
pressure ${\cal P}(\mu)$) into the Tolman-Oppenheimer-Volkoff equations for relativistic
stellar structure. Indeed, such a programme is a necessary prerequisite for
determining whether postulated ground states such as color-superconducting
quark matter
could ever exist in our universe.

The current consensus~\cite{Halasz:1998qr} is that as baryon density $n$ increases, the state best
described as ``nuclear matter'', {\it viz\/}. a degenerate system of neutrons and
protons with $n\simeq0.45$fm$^{-3}$ which is the favoured ground state once
$\mu$ exceeds its {\it onset} value $\mu_o\simeq924$MeV, is somehow succeeded
by an alternative degenerate system called ``quark matter''. The properties of
quark matter have been the subject of intense speculation over the past decade;
it has been suggested that for sufficiently low temperatures $T$, as a result of
quark Cooper pair condensation at the Fermi surface, global, local and even
translational symmetries may be spontaneously broken in the ground state,
leading to exotic phenomena such as color superconductivity (CSC) or even
crystallization~\cite{Alford:2007xm}. From a
theoretical standpoint these scenarios are most readily studied at weak
coupling, which in the thermodynamic limit 
can only be quantitatively accurate at asymptotically high
densities where $\mu\gg\Lambda_{QCD}$, and yet the passage from nucleons to
quarks referred to above clearly requires a non-perturbative treatment. At the
very least, a reliable matching between perturbative and non-perturbative regimes is
required. A recent calculation to this end has appeared in \cite{Kurkela:2009gj}. Other
theoretical issues which naturally arise in this context include: what precisely
is meant by ``degenerate matter''? (this is usually taken to mean a system with a
well-defined Fermi surface characterized by a momentum scale $k_F$, but this
definition is not gauge-invariant), and, does chiral symmetry
restoration and/or deconfinement occur as $\mu$ increases? And, if so, to what
extent do the transitions resemble those known to occur for $\mu=0$ as $T$ is
raised?

An important reason why a non-perturbative understanding of QCD with $\mu\not=0$
has not progressed as much as in other areas is the unavailability of lattice
gauge theory simulations performed using standard techniques \cite{Hands:2007by}. 
A system with $\mu\not=0$ is not invariant under time
reversal, since a bias is introduced in favour of particle propagation in the
positive $t$ direction. In the Euclidean metric this results in asymmetry between
$i$ and $-i$, so that for instance the Polyakov line defined by $P(\vec
x)\equiv\Tr\Pi_{t=1}^{N_t}U_0(\vec x,t)$ has the property $\langle
P\rangle\not=\langle P^\dagger\rangle^*$, implying that in a medium with $n>0$ the
free energy of a static color source differs from that of an anti-source.
Crucially, it implies that in general the Euclidean action is complex-valued,
and hence has a fluctuating phase $\phi = \Im S$. Since the functional measure $e^{-S}$ is no
longer positive definite, the Monte Carlo importance sampling used in lattice
simulations is inoperable. Indeed, it appears that physically acceptable results
can only be obtained if delicate cancellations are correctly handled over a much
larger region of configuration space than that normally
considered~\cite{Cohen:2003kd,Osborn:2004rf}.

Attempts to evade this so-called {\it Sign Problem\/} fall into three classes.
First, one can run simulations at $\mu=0$ and attempt to calculate operator
expectation values via {\it re-weighting}
\begin{equation}
\langle{\cal O}\rangle=\frac{\langle\langle{\cal
O}e^{i\phi}\rangle\rangle}{\langle\langle e^{i\phi}\rangle\rangle},
\end{equation}
where $\langle\langle\ldots\rangle\rangle$ denotes averaging with respect to a
suitably chosen real measure. This approach has been found to be particularly
effective in the vicinity of the quark-hadron phase transition for
$\frac\mu T\lapprox1$~\cite{Fodor:2001pe}, 
but must fail in the thermodynamic limit since the ratio of
two partition functions $\langle\langle e^{i\phi}\rangle\rangle\sim e^{-CV}$,
resulting in a disastrously poor signal to noise ratio as $V\to\infty$. Other
methods which can be applied in this physical regime, relevant for the hot
medium produced in RHIC collisions, rely on analytic continuation of results
generated from simulations with a real action, either at $\mu=0$ by calculating
successive terms in a Taylor expansion~\cite{Allton:2002zi}, 
or by simulating with imaginary chemical
potential (corresponding to real constant abelian electrostatic
potential)~\cite{deForcrand:2002ci}.
These latter approaches work in the thermodynamic limit, but are necessarily
limited by a finite radius of convergence.  The three methods have achieved some
consensus in mapping out the quark-hadron transition line and determining the
equation of state for small $\frac\mu T$; whether they will ultimately prove capable
of saying something about a possible critical point in the $(\mu,T)$ plane is as
yet unresolved.

A second approach is to study gauge theories without a sign problem,
{\it ie\/}. 
where the functional measure
remains positive definite for $\mu\not=0$. These include QCD with isospin
chemical potential ({\it ie\/}. with $\mu_I\equiv\mu_d=-\mu_u$)~\cite{Son:2000xc}, 
and theories with real matter
representations such as the fundamental of $SU(2)$ (or the {\bf6} of
$SU(4)$) with
$N_f$ even, or any theory with adjoint quarks~\cite{Kogut:2000ek}. 
The generic feature of such
models is a degeneracy between mesons and baryons at $\mu=0$: hadron multiplets
contain both mesonic $q\bar q$ states and $qq$ or $\bar q\bar q$ states which
carry a non-zero baryon charge (in the case of $\mu_I\not=0$ the latter role is
played by $\pi$-mesons with non-zero $I_3$). In such theories the onset takes place for
$\mu_o\sim M_\pi$, where $M_\pi$ is the mass of the pseudo-Goldstone boson
associated with chiral symmetry breaking at $\mu=T=0$. Theories with
sufficiently small quark mass so there is  a large separation between Goldstone
and hadronic scales are then well-described by chiral perturbation theory
($\chi$PT), in which only Goldstone degrees of freedom, some baryonic, are
retained. For $\mu\gapprox\mu_o$ the resulting system is an (arbitrarily) dilute
Bose-Einstein condensate (BEC) formed from tightly-bound $qq$
bosons~\cite{Kogut:2000ek}. These
models therefore fail to describe nuclear matter, but may still have something
important to tell us about quark matter at higher densities.  As $\mu$ is
increased beyond onset there is a smooth rotation from the chiral condensate
characteristic of the vacuum at $T=\mu=0$ to a gauge invariant diquark
condensate $\langle qq\rangle\not=0$.  Since this condensate breaks a global,
rather than a local, symmetry, the ground state is superfluid but is not a color superconductor.
The predictions of $\chi$PT for $\mu\gapprox\mu_o$ have been quantitatively
confirmed by several lattice simulations~\cite{Hands:2000ei}; more recent lattice works have
explored the breakdown of $\chi$PT at larger values
$\mu\sim\Lambda_{QCD}$~\cite{Hands:2006ve}. A related study of $\rho$-meson
condensation in large-$N_c$ QCD with $\mu_I\not=0$ uses the methods of AdS/CFT
duality~\cite{Aharony:2007uu}.

We should mention two more radical approaches to simulating theories with
$\mu\not=0$. It is possible to mitigate or even eliminate a sign problem by
transforming to a different set of field variables via, {\it
  eg.\/}~the 
exploitation of a duality symmetry~\cite{Hands:2007by}. 
Two recent papers have applied this idea to convert bosonic
field theories in three dimensions to loop gases, which are then
simulable~\cite{Endres:2006xu,Banerjee:2010kc}. A related treatment of nuclear matter in the strong-coupling limit of lattice QCD has also appeared in \cite{deForcrand:2009dh}.
Finally, there is the possibility of generating representative field
configurations by integrating a stochastic differential equation, the {\it
Langevin} equation, in which the complex drift terms resulting from the action
force the field variables to evolve in an extended, complexified space in which
the large regions where observables are swamped by phase fluctuations can be
avoided~\cite{Aarts:2008rr}. 
Again, this method has been successfully applied in certain bosonic
cases~\cite{Aarts:2008wh}.

In our approach, the fact that the action is complex for $\mu\neq0$
turns out not to be a problem but it does have important
consequences. In the small volume theory, the
effective action is defined over the eigenvalues of the 
the Polyakov line $P$, which is a unitary matrix whose eigenvalues 
can be written
$e^{i\theta_i}$, $i=1,\ldots,N$. The functional integral reduces to an
integral over the angles $\{\theta_i\}$.
At large $N$ the
functional integral is dominated by a single saddle-point but since
the action is not real this saddle-point configuration lies out in
the complex plane where the $\theta_i$ are no longer real. As
a consequence $\langle P\rangle\neq\langle
P^\dagger\rangle^*$. 

In the following sections we will summarize the derivation of the
action of QCD formulated on $S^1 \times S^3$ from one loop
perturbation theory. Then we will present calculations of several
observables as a function of the chemical potential, working at low
temperatures, in both the small $N$, and the large $N$ limits. For
sufficiently small $N$ it is possible to calculate the partition
function, and thus any observable derivable from it, by simply
numerically performing the integrals over the gauge fields. 
Of course for finite $N$ in finite volume there are no sharp phase
transitions; nevertheless, they show up qualitatively in the behaviour
of observables. We present
results for $N = 3$ which involves performing integrals over the 2
eigenvalues of the Polyakov line which is an $SU(3)$ matrix.
In this exploratory
study we consider $N_f = 1$ Dirac fermion flavor, and consider the
limit of a light quark $m R = 0$, and a heavy quark. What
we find is that the fermion number, pressure, and energy rise in
discrete levels as a function of the chemical potential. This is
reminiscent of the quantum hall effect of QED \cite{Persson:1994pz,Zeitlin:1994vs} where in our
case the discrete levels result due to restricting our calculation to
small spatial volumes. The magnetic field $B$ which causes the quantum
hall effect in QED is loosely analogous to our $R^{-2}$. The analogy
can be taken a bit further in that in both QED with non-zero external
magnetic field, and QCD in a small spatial volume both exhibit the
level structure of the fermion number as a function of the chemical potential only in the
low temperature limit. Increasing the temperature causes the levels to
become smoothed out. Or, when $B \sim R^{-2}$ is small (large volumes)
the levels are also smoothed out. 

In the large $N$ limit it is necessary to consider the gauge field,
corresponding to the angles of the Polyakov line, as a
distribution on a contour. From the
equation-of-motion the saddle-point distribution of the Polyakov line
eigenvalues can be calculated analytically and plotted by mapping the
angles from an arc on the unit circle to a contour over the same range
of angles in the complex plane. What we observe is in agreement with
the finite $N$ results. The contour on which the Polyakov line
eigenvalues are distributed is closed, corresponding to the confined phase, in between
level transitions, and opens up while the transitions between levels
are taking place. This is the characteristic feature of a third order,
Gross-Witten transition [in the Ehrenfest classification], and indeed the third derivative of the grand
potential is discontinuous at each level crossing \cite{Gross:1980he,Wadia:1979vk,Wadia:1980cp}.

\section{Background}

Here we will summarize the one-loop formulation of QCD on $S^1 \times S^3$ which was derived for $SU(N)$ gauge theories with more general matter content in the beautiful paper \cite{Aharony:2003sx}. This section does not present new material and is merely included for completeness. The partition function of QCD at finite temperature $T = 1/ \beta$, for $N_f$ quark flavours, each with a mass $m_f$ and coupled to a chemical potential $\mu_f$ is given in Euclidean space by
\EQ{
Z_{QCD} = \int {\cal D} A {\cal D} {\bar \psi} {\cal D} \psi e^{- \int_0^{\beta} d \tau \int d^3 {\boldsymbol x} \LAG_{QCD}}
}
where $\psi$ and ${\bar \psi}$ are the fundamental and anti-fundamental fermion fields, respectively, and $A$ is the $SU(N)$ gauge field, $A_{\mu} = A_{\mu}^a T^a$. The Lagrangian is
\EQ{
\LAG_{QCD} = \frac{1}{4 g^2} {\rm Tr}_F \left( F_{\mu \nu} F_{\mu \nu} \right) + \sum_{f=1}^{N_f} {\bar \psi}_f \left( \Dsl_F (A) - \gamma_0 \mu_f + m_f \right) \psi_f ,
}
with covariant derivative
\EQ{
D_{\mu} (A) \equiv \partial_{\mu} - A_{\mu} ,
}
and field tensor
\EQ{
F_{\mu \nu} \equiv \left[D_{\mu} (A), D_{\nu} (A) \right] = \partial_{\nu} A_{\mu} - \partial_{\mu} A_{\nu} + \left[A_{\mu}, A_{\nu}\right] .
}
When the spacetime geometry is $S^1\times S^3$, only the component
$A_0$ along the $S^1$ has a zero mode. The idea is to construct a Wilsonian
effective action for this mode. To this end we 
decompose $A_0 = \alpha + g {\mathscr A}_0$ where,
without-loss-of-generality, we can choose the background field $\alpha$
to consist of the diagonal elements of $A_0$ while the fluctuation 
$g{\mathscr A}_0$
consists of the off-diagonal elements. The background field breaks the
gauge symmetry from $SU(N)$ to its maximal abelian group $U(1)^{N-1}$
and gives mass to the off-diagonal modes which can then be integrated out.
Gauge
fixing with Feynman gauge and retaining the one-loop contributions
puts the Lagrangian in the form 
\EQ{
\begin{aligned}
\LAG_{QCD} = &- \frac{1}{2} {\mathscr A}_{0}^a ( D_0^2 (\alpha) + \Delta^{(v)} ) {\mathscr A}_{0}^a - \frac{1}{2} A_{i}^a ( D_0^2 (\alpha) + \Delta^{(v)} ) A_i^{a}\\
&- {\bar c} ( D_0^2 (\alpha) + \Delta^{(s)} ) c + {\bar \psi} ( \Dsl_F (\alpha) - \gamma_0 {\cal M} + M ) \psi ,
\end{aligned}
}
where $\psi$ (${\bar \psi}$) is an $N_f$ component vector containing the $\psi_f$ (${\bar \psi}_f$), $M$ (${\cal M}$) is an $N_f \times N_f$ diagonal matrix containing the $m_f$ ($\mu_f$) as the diagonal elements, ${\bar c}$ and $c$ are complex Grassmann-valued ghost fields resulting from gauge fixing. Here $\Delta^{(s)}$ and $\Delta^{(v)}$ represent the scalar and vector Laplacians, respectively, where $\Delta^{(s)} = g^{-1/2} \partial_{\mu} (g^{1/2} \partial_{\mu})$ and $\Delta^{(v)} A^{i} = \nabla_j \nabla^{j} A^{i} - R^{i}_{j} A^{j}$ with $R_{i j}$ the Ricci tensor of $S^3$. It is useful to decompose the spatial gauge field as $A_i = B_i + C_i$, where $B_i$ is the transverse (T) component with $\nabla_i B_i = 0$, and $C_i$ is the longitudinal (L) component with $C_i = \nabla_i f$. Then we have
\EQ{
\begin{aligned}
\LAG_{QCD} = &- \frac{1}{2} {\mathscr A}_{0}^a ( D_0^2 (\alpha) + \Delta^{(s)} ) {\mathscr A}_{0}^a - \frac{1}{2} B_{i}^a ( D_0^2 (\alpha) + \Delta^{(v,T)} ) B_i^{a}\\
&- \frac{1}{2} C_{i}^a ( D_0^2 (\alpha) + \Delta^{(v,L)} ) C_i^{a} - {\bar c} ( D_0^2 (\alpha) + \Delta^{(s)} ) c + {\bar \psi} ( \Dsl_F (\alpha) - \gamma_0 {\cal M} + M ) \psi .
\end{aligned}
}
Performing the Gaussian integrals we obtain a simple form for the effective partition function (there is yet the integral over $\alpha$ to perform to obtain the full partition function),
\EQ{
Z (\alpha) = {\det}_{\ell=0}^{1/2} \left( -D_0^2 (\alpha) - \Delta^{(s)} \right) {\det}^{-1} \left( - D_0^2 (\alpha) - \Delta^{(v,T)} \right) \det \left( \Dsl_F (\alpha) - \gamma_0 {\cal M} + M \right) ,
\label{simpleZ}
}
where we note that on $S^3$, the eigenfunctions of the scalar
Laplacian have energies $\varepsilon_{\ell}^{(s)}$ and degeneracies $d_l^{(s)}$ given by
\SP{
\Delta^{(s)} Y_{\ell}(\hat\Omega) &= - \varepsilon_\ell^{(s)
  2} 
Y_{\ell}(\hat\Omega)\ ,\\
\varepsilon_\ell^{(s) 2}&= \ell(\ell+2) R^{-2}\ ,\\
d_\ell^{(s)} &= (\ell+1)^2\ ,
}
where $\ell=0,1,2\ldots$ and $R$ is the radius of $S^3$.
To obtain (\ref{simpleZ}) we used the fact that the vector Laplacian acts on the longitudinal vectors $C_i$ as
\EQ{
\Delta^{(v)} ( \nabla_k f ) = \left( \nabla^i \nabla_i \delta^j_k - R^j_k \right) \nabla_j f = \nabla_k \left( \nabla^i \nabla_i f \right)
}
which results in the same spectrum as for scalars, $\Delta^{(v,L)} C_i =
\Delta^{(s)} C_i$, with the exception that $\ell\geq1$ for
vector fields.\footnote{The derivative of the $\ell=0$ mode vanishes
  since it is a constant.} This leads to the almost-cancellation between the
${\mathscr A}_0$, $C_i$, ${\bar c}$ and $c$ terms, with the $\ell = 0$
contribution, ${\det}_{\ell=0}^{1/2} \left( -D_0^2 (\alpha) - \Delta^{(s)}
\right)$, as the only remaining piece. 

To evaluate the fermion contribution we need the identity
\EQ{
\gamma^i \gamma^j \nabla_i \nabla_j = g^{i j} \nabla_i \nabla_j - \frac{1}{4} {\cal R} .
}
where ${\cal R}$ is the scalar curvature of $S^3$. Then evaluation of the fermion determinant proceeds as follows performing the determinants over the flavor then spinor degrees of freedom,
\EQ{
\begin{aligned}
\log Z_f (\alpha) &= \log \det ( \Dsl_F (\alpha) - \gamma_0 {\cal M} + M )\\
&= \frac{1}{2} \log \det \left[ - \left( \Dsl_F (\alpha) - \gamma_0 \mu \right)^2 + m^2 \right]^{N_f}\\
&= 2 N_f \log \det \left[ - \left( D_0 (\alpha) - \mu \right)^2 - \Delta^{(f)} + \frac{1}{4} {\cal R} + m^2 \right] .
\end{aligned}
}
The eigenvalues and degeneracies of the spinor Laplacian on $S^3$ are given by
\SP{
\left( \Delta^{(f)} - \frac{1}{4} {\cal R} \right) \psi &= -
\varepsilon_\ell^{(f) 2} \psi\ ,\\
\varepsilon_\ell^{(f) 2} &= \left( \ell + \tfrac{1}{2} \right)^2 R^{-2}\ ,\\
d_\ell^{(f)} &= \ell \left( \ell + 1 \right) ,
}
where $\ell=1,2,\ldots$.
Regarding the $S^1$ around which the fermions have anti-periodic
(thermal) boundary conditions, 
the eigenvalues of the operator $D_0 (\alpha)$ are discretized in
terms of the Matsubara frequencies, $\omega_n = (2 n + 1) \pi /
\beta$, and given by
\EQ{
D_0 (\alpha) \rightarrow i \omega_n - \alpha^a T^a .
}
The fermion contribution then takes the form
\SP{
\log Z_f (\alpha)
&= 2 N_f \Tr_R \sum_{n \in {\boldsymbol Z}} 
\sum_{\ell =
  1}^{\infty} d_\ell^{(f)} \left[\log\left(
\omega_n^2
+ (
    \varepsilon^{(f,m)}_\ell - \mu-\alpha )^2 
\right)\right.\\ &~~~~~~~~~~~~~~\left. + \log \left( \omega_n^2 + (
    \varepsilon^{(f,m)}_\ell + \mu+\alpha )^2 \right) \right]\ ,
}
where $\varepsilon_\ell^{(f,m)} 
= \sqrt{\varepsilon_\ell^{(f) 2} + m^2}$.

We define the Polyakov line order parameter, $P$, by the path-ordered exponential of the temporal gauge field. In terms of the constant temporal background field $\alpha \equiv i \theta / \beta$, it is
\EQ{
P = {\mathscr P} e^{\int_0^{\beta} dt\, A_0 (x)} = e^{\beta \alpha} = e^{i \theta} = \text{diag}\big(e^{i \theta_1},
... , e^{i \theta_N}\big)\ .
}
Then, following \cite{Gross:1980br},
\EQ{
\log Z_f (\theta_i) = - 2 N_f \Tr_R \sum_{l = 1}^{\infty} d_\ell^{(f)}
\sum_{n = 1}^{\infty} \frac{(-1)^n}{n} e^{- n \beta
  \varepsilon^{(f,m)}_\ell}\sum_{i=1}^N \left[ e^{n \beta \mu + i n
    \theta_i} + e^{- n \beta \mu - i n \theta_i} \right] , 
}
where we have dropped terms that are independent of the gauge field as their contribution cancels when taking expectation values.

We follow a similar procedure to obtain the boson contribution. Given the eigenvalues and degeneracies of the transverse vector Laplacian on $S^3$,
\EQ{
\Delta^{(v)} B^i_{\ell}(\hat\Omega)&= - \varepsilon_\ell^{(v,T) 2}
B^i_{\ell}
(\hat\Omega)\ ,\\
\varepsilon_\ell^{(v,T) 2}&= (\ell+1)^2R^{-2}\ ,\\
d_\ell^{(v,T)}& = 2\ell(\ell+2)\ ,
}
for $\ell=1,2,\ldots$, we have
\EQ{
\begin{aligned}
&\log Z_b (\theta_i)
= \frac{1}{2} \log {\det}_{\ell=0} \left( - D_0^2 (\alpha) - \Delta^{(s)} \right) - \log \det \left( - D_0^2 (\alpha) - \Delta^{(v,T)} \right)\\
&= \sum_{n = 1}^{\infty} \frac{1}{n} \left( - 1 +
  \sum_{\ell = 1}^{\infty} d_\ell^{(v,T)} e^{- n \beta
    \varepsilon_\ell^{(v,T)}} \right)
\sum_{ij=1}^N\cos (n(\theta_i-\theta_j)) ,
\end{aligned}
}
where we used the fact that the trace in the adjoint representation
\EQ{
\Tr_A(P)=\sum_{ij=1}^N\cos (n(\theta_i-\theta_j))\ .
}
Adding the boson and fermion contribution the total one-loop effective action is
\SP{
S(\theta_i) = &\sum_{n=1}^{\infty} \frac{1}{n} \left( 1 - z_b (n \beta/R) \right)\sum_{i,j=1}^N\cos(n(\theta_i-\theta_j))\\
&+ \sum_{n=1}^{\infty} \frac{(- 1)^n}{n} N_f z_f (n \beta/R, m
R)\sum_{i=1}^N 
\left[ e^{n \beta \mu+in\theta_i} + 
e^{-n \beta \mu-in\theta_i} \right] ,
\label{pop}
}
where we have defined
\SP{
z_b(\beta/R)&=\sum_{\ell=1}^\infty
d_\ell^{(v,T)}e^{-\beta\varepsilon_\ell^{(v,T)}}\\ 
&=2\sum_{\ell=1}^\infty\ell(\ell+2)e^{-\beta(\ell+1)/R}\\
&=\frac{6e^{-2\beta/R}-2e^{-3\beta/R}}{(1-e^{-\beta/R})^3}\ ,
\label{jhh}
}
and
\SP{
z_f(\beta/R,mR)&=\sum_{\ell=1}^\infty
d_\ell^{(f)}e^{-\beta\varepsilon_\ell^{(f,m)}}\\
&=2\sum_{\ell=1}^\infty\ell(\ell+1)e^{-\beta 
\sqrt{(\ell+\frac12)^2+m^2R^2}/R}\\
&=\frac{2m^2R^3}{\beta}K_2(\beta m)-\frac{mR}2K_1(\beta m)\\
&~~~~~~~~~~~~~~~~+
4\int_{mR}^\infty dx\,\frac{x^2+\tfrac14}{e^{2\pi
    x}+1}\sin(\beta \sqrt{x^2-m^2R^2}/R)\ .
}

Notice that the first term in the sum \eqref{pop}, up to an 
unimportant constant,
can also be interpreted as the Vandermonde Jacobian
contribution resulting when converting from an integral over unitary
matrices of the form $P = e^{i \theta}$ to an integral over the
eigenvalues, $\theta_i$, of $\theta$. That is, 
\EQ{
\begin{aligned}
\int dP &= \int \prod_{i=1}^N d\theta_i \prod_{j<i}^{N} \sin^2 \left( \frac{\theta_i - \theta_j}{2} \right)\\
&= \int \prod_{i=1}^{N} d \theta_i \exp \left[ -S_{Vdm} \right] ,
\end{aligned}
}
In the low temperature limit ($\beta \rightarrow \infty$) we
have $z_b(\infty)=0$
and so the gluonic contribution to the action reduces to the Vandermonde piece, $S_\text{Vdm}$, defined above:
\EQ{
S = S_\text{Vdm} + S_f
}
and so in this limit the theory reduces to that to that of an $N\times
N$ unitary matrix model
\EQ{
Z=\int dP\,e^{-N\,\Tr V(P)}\ ,
\label{mmod}
}
with a potential determined by the quark contributions that we can
write as
\EQ{
V(P)=-\sum_{\ell=1}^\infty
\sigma_\ell\left[\log(1+e^{\beta(\mu-\varepsilon_\ell)}P)+
\log(1+e^{\beta(-\mu-\varepsilon_\ell)}P^\dagger)\right]\ .
}
with the definitions that we use from now on
\EQ{
\sigma_\ell=2\ell(\ell+1)\frac{N_f}N\ ,~~~~\varepsilon_\ell\equiv
\varepsilon_\ell^{(f,m)}=\sqrt{
m^2+(\ell+\tfrac12)^2R^{-2}}\ .
}

\section{Finite $N$: $N = 3$}

Even though there are no sharply-defined phase transitions for a finite $N$ theory valid in a small spatial volume, taking the low temperature (large $S^1$) limit is not
that far removed from a limit where well-defined phase transitions are possible: as the temperature is decreased
transitions as a function of the chemical potential appear more and more sharp, even if on a microscopic scale
they are always continuous. To develop an understanding of the physics
at non-zero chemical potential we calculate several observables
from the low temperature partition function,

\EQ{
\begin{aligned}
Z (\beta/R) &= \int \left[ {\mathrm d} \theta \right] \text{exp}\left[ - \sum_{n=1}^{\infty} \frac{1}{n} \left[ \Tr_A (P^n) + (-1)^n N_f z_f (n \beta/R, m R) e^{n \beta \mu} \Tr_F (P^n) \right] \right] ,
\end{aligned}
}
where $[d \theta] = \prod_{i=1}^{N} d \theta_i$. These include:


\begin{eqnarray}
\text{Fermion number:} & \hspace{4mm} {\mathscr N}_i &= \frac{1}{\beta} \left( \frac{\partial \log Z}{\partial \mu_i} \right)\\
\text{Polyakov lines:} & \hspace{4mm} \PP_1 &= \frac{1}{Z} \int \left[ {\mathrm d} \theta \right] e^{-S} \left( \sum_{i=1}^{N} e^{i \theta_i} \right)\\
& \hspace{4mm} \PP_{-1} &= \frac{1}{Z} \int \left[ {\mathrm d} \theta \right] e^{-S} \left( \sum_{i=1}^{N} e^{-i \theta_i} \right)\\
\text{Pressure:} & \hspace{4mm} {\cal P} & = \frac{1}{\beta} \left( \frac{\partial \log Z}{\partial V_3} \right)\\
\text{Energy:} & \hspace{4mm} E &= - {\cal P} V_3 + {\mu}_i \, {\mathscr N}_i \\
\text{Chiral condensate:} & \hspace{5mm} \langle {\bar \psi} \psi \rangle & = - \frac{1}{\beta V_3} \lim_{m \rightarrow 0} \left( \frac{\partial \log Z}{\partial m} \right)\\
\text{Average phase:} & \hspace{4mm} \langle e^{i \phi} \rangle_{p q} &= \frac{Z}{Z_{p q}}
\end{eqnarray}

\noindent where the $p q$ in the average phase refers to the phase-quenched theory to be discussed later. In what follows we first calculate each of these observables for arbitrary $N$, not performing the integrals over $\theta_i$. Then we present results for $N = 3$ by integrating numerically over the $\theta_i$. Each observable is calculated as an expectation value with the form
\EQ{
{\cal O} \equiv \frac{\int \left[ {\mathrm d} \theta \right] e^{-S} {\cal O}}{\int \left[ {\mathrm d} \theta \right] e^{-S}} \xrightarrow[N=3]{} \frac{\int {\mathrm d}\theta_1{\mathrm d}\theta_2 e^{-S} {\cal O}}{\int {\mathrm d}\theta_1{\mathrm d}\theta_2 e^{-S}}
}
where the integrals over $\theta_3$ drop out as $\theta_3 = -\theta_1 - \theta_2$ by the $SU(N)$ condition. These results are multiplied by factors of $\beta$ and / or $V_3$ as needed to make them dimensionless. In this paper we present results for $N_f = 1$ Dirac fermion flavor. We calculate the above observables considering first the case of a massless quark, and then the case of a quark with large mass.

\subsubsection{Fermion number ${\mathscr N}$ for $m = 0$}


\begin{figure}[t]
\center
\includegraphics[width=9cm]{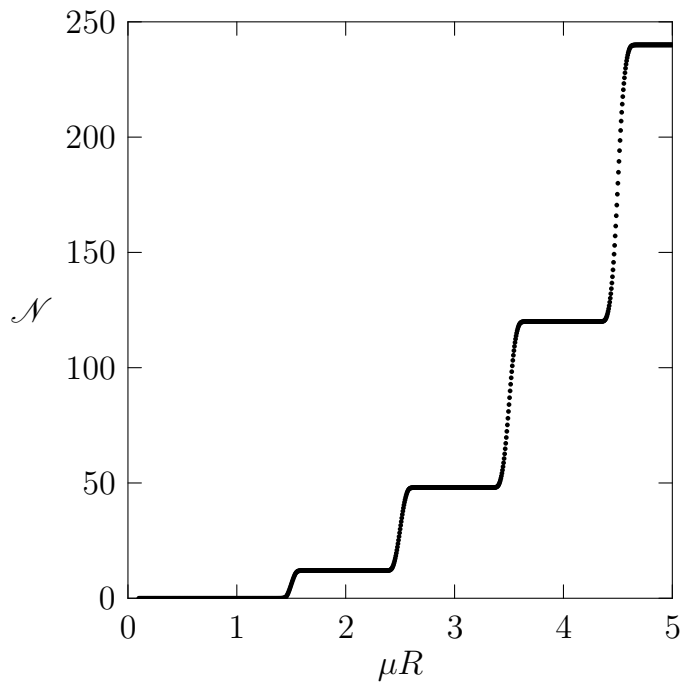}
\caption{Expectation value of the fermion number as a function of the quark chemical potential for QCD on $S^1 \times S^3$. $N=3$, $N_f = 1$, $m = 0$, $\beta / R = 30$ (low $T$).}
\label{fermnum}
\end{figure}


The fermion number ${\mathscr N}$ gives the number of quarks minus the number of
antiquarks in the volume of $S^3$, $V_3 = 2 \pi^2 R^3$, where $R$ is the radius of $S^3$. From Figure
\ref{fermnum}, which shows ${\mathscr N}$ as a function of $\mu R$ for low temperatures for a single massless quark flavor, an occupation level structure is apparent. The fermion number in this limit is 
\EQ{
\begin{aligned}
{\mathscr N} &= \frac{1}{\beta} \left( \frac{\partial \log Z}{\partial \mu} \right)\\
&= \frac{-1}{\beta Z} \int \left[ {\mathrm d} \theta \right] e^{-S} \left( \frac{\partial S}{\partial \mu} \right)\\
&\xrightarrow[\beta \rightarrow \infty]{} \frac{N_f}{Z} \int \left[
  {\mathrm d} \theta \right] e^{-S} \sum_{\ell=1}^{\infty}
\sum_{i=1}^{N} 2 \ell (\ell+1) \left[ \frac{e^{\beta \mu}}{e^{\beta
      \mu} + e^{-i \theta_i + \beta (\ell+\frac12) / R}} \right] , 
\end{aligned}
} 
where the derivative of the action with respect to the chemical
potential brings down a factor of $n \beta$, leading to a geometric
series which gets summed to give the Fermi-Dirac distribution. Ignoring the Polyakov line for the moment, it is clear from the general form of the Fermi-Dirac distribution function,
\EQ{
f(\varepsilon_\ell) = \frac{1}{1+e^{\beta (\varepsilon_\ell - \mu)}},
}
that the transitions occur when $\varepsilon_\ell - \mu$ changes sign, {\it i.e.\/}, when $\mu$ passes an energy level. When $\mu \ll \varepsilon_l$ then $f \sim 0$. When $\mu \gg \varepsilon_\ell$ then $f \sim 1$. The system is in transition for $\left| \beta (\varepsilon_\ell - \mu) \right|$ small. This
shows that each level $L$ has a net number of quarks given by 
\EQ{
{\mathscr N}_L = N N_f \sum_{\ell=1}^{L} 2 \ell (\ell+1)\ .
}
Each new level starts at
\EQ{
(\mu R)_0 = L+\tfrac12\ ,
}
has a level width
\EQ{
(\Delta \mu R)_{\Delta l} = 1\ ,
}
and a transition width given by the width of the Fermi-Dirac distribution function. This goes like the temperature,
\EQ{
(\Delta \mu R)_L \sim \frac{R}{\beta} .
}

Since the level width in $\mu$ goes like $1/R$, then the levels should become
narrower with increasing spatial volume when considered as a function
of the chemical potential alone. In addition, as we will show in the next
section, the width of the levels also depends on the quark mass $m$,
and taking $m$ large also causes the steps to become narrower, at
least until $\mu \gg m$. For large values of $\mu$ or small values of $\beta$ (finite temperature)
the {\it transition} width is larger. 

It is interesting to compare with the results of
\cite{Banerjee:2010kc}. The authors observe the same level structure
in the particle number in the non-linear $O(2)$ sigma model. Their
results for the particle number as a function of the chemical potential
(not multiplied by the spatial extent) show that the levels become narrower as
the spatial volume is increased, and the particle number appears more
continuous. Our results indicate a level width $\Delta \mu =
\frac{1}{R}$ (in the massless limit) and are thus qualitatively
consistent with theirs. 

\subsubsection{Polyakov lines: $\PP_1$ and $\PP_{-1}$ for $m = 0$}


\begin{figure}[t]
\center
\includegraphics[width=9cm]{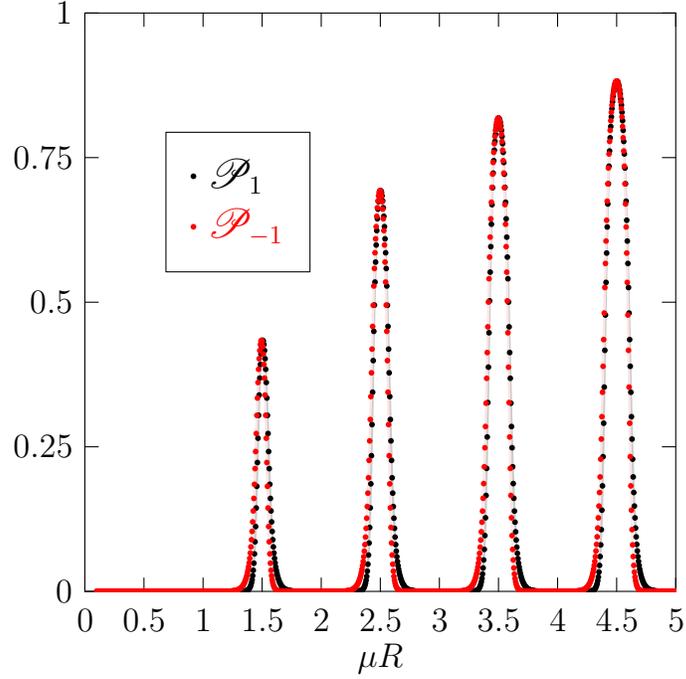}
\caption{Expectation values of Polyakov lines $\PP_1$ and $\PP_{-1}$ as a function of the chemical potential for $N = 3$, $N_f = 1$, $m=0$, $\beta / R = 30$ (low $T$).}
\label{polys}
\end{figure}

\begin{figure}[t]
  \hfill
  \begin{minipage}[t]{.49\textwidth}
    \begin{center}
\includegraphics[width=0.99\textwidth]{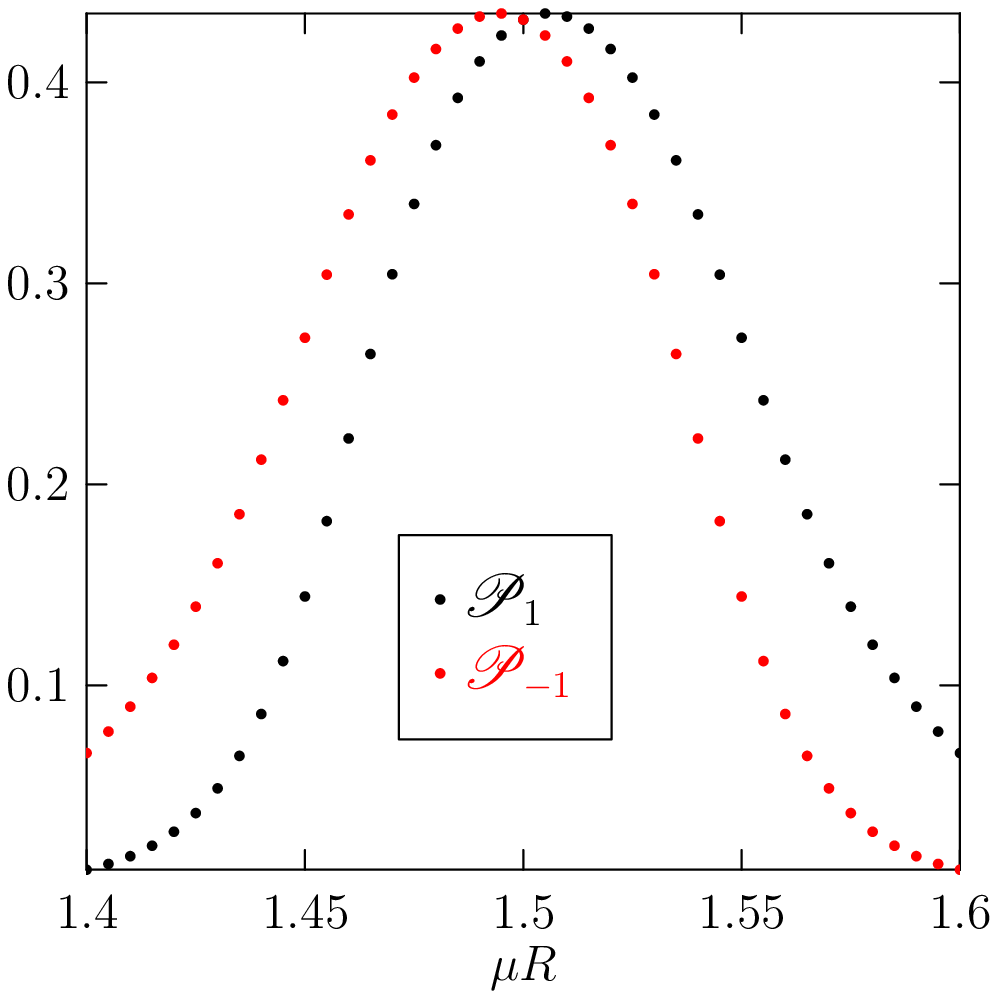}
    \end{center}
  \end{minipage}
  \hfill
  \begin{minipage}[t]{.49\textwidth}
    \begin{center}
\includegraphics[width=0.99\textwidth]{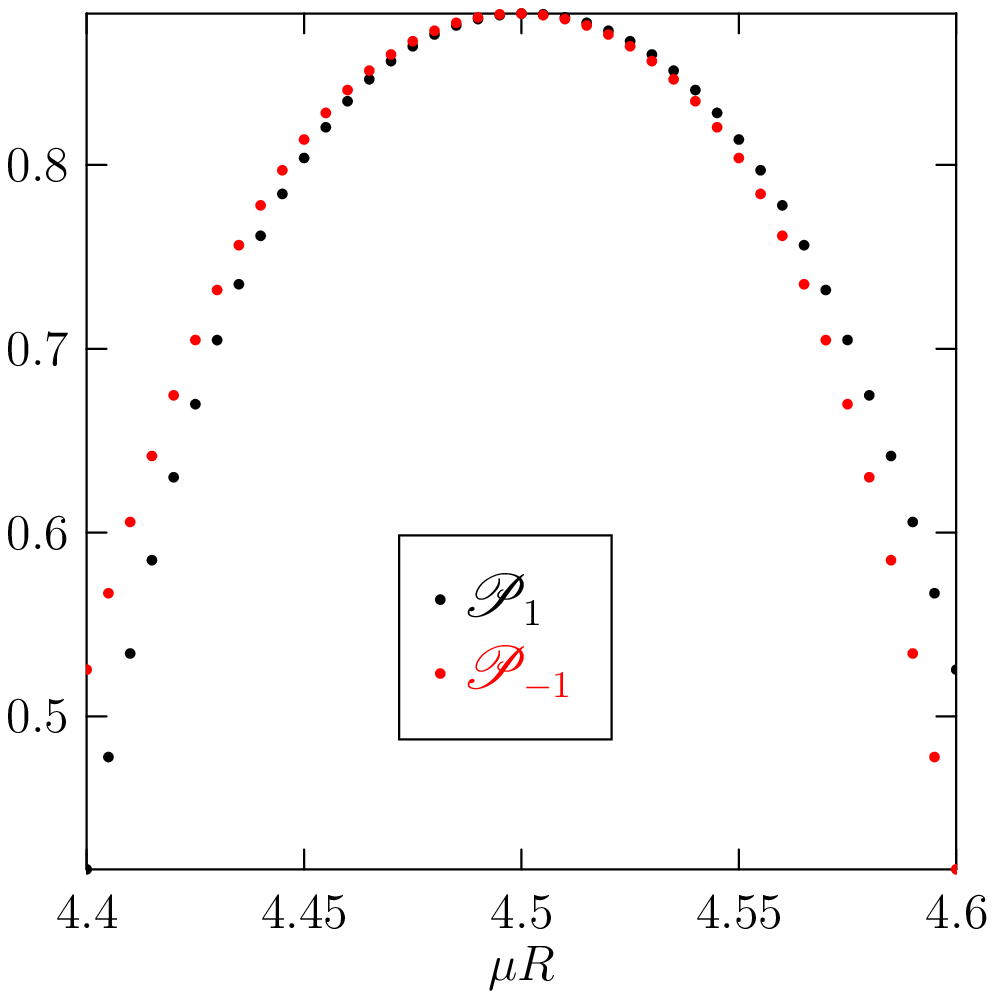}
    \end{center}
  \end{minipage}
  \hfill
\caption{$\PP_1$ and $\PP_{-1}$ as a function of the chemical potential at the first
  transition (Left), and the fourth (Right). $N = 3$, $N_f = 1$, $m =
  0$, $\beta / R = 30$ (low $T$). The width of the deconfined regions increases with $\mu R$.} 
\label{polys_close}
\end{figure}

At zero chemical potential, $\PP_1 = \langle \Tr P \rangle$ and $\PP_{-1} = \langle \Tr P^{\dagger} \rangle$ are complex conjugates, but for non-zero chemical potential this is no longer true\footnote{The matrix model studied in \cite{Dumitru:2005ng} has a similar fermion term and also shows this effect.}. The Polyakov line expectation values are
\EQ{
\PP_1 = \frac{1}{Z} \int \left[ {\mathrm d} \theta \right] e^{-S} \,
\sum_{i=1}^{N} e^{i \theta_i}  , 
}
and
\EQ{
\PP_{-1} = \frac{1}{Z} \int \left[ {\mathrm d} \theta \right] e^{-S}
\, \sum_{i=1}^{N} e^{- i \theta_i}. 
} 
Figure \ref{polys} shows $\PP_1$ and $\PP_{-1}$ as a function of $\mu
R$. Each spike in $\PP_1$ and $\PP_{-1}$ corresponds to a level transition
in ${\mathscr N}$. Even though their behaviour as a function of $\mu R$ is
similar, the peaks of $\PP_{-1}$ always preceed those of $\PP_{1}$ at the
start and finish of each level transition.

Taking the temperature lower than that of Figure \ref{polys} causes the peaks to be narrower while maintaining the same height. Taking $T \rightarrow 0$ would make the peaks appear as infinitely narrow lines, occurring precisely at $\mu R = 1.5, 2.5, ...$.

In Figure \ref{polys_close} we compare the first and fourth
transitions. For non-zero $T$, as $\mu R$ increases the transition width, and thus the width of the deconfined regions,
increases. 

\subsubsection{Pressure ${\cal P}$ for $m = 0$}

\begin{figure}[t]
  \hfill
  \begin{minipage}[t]{.49\textwidth}
    \begin{center}
\includegraphics[width=0.99\textwidth]{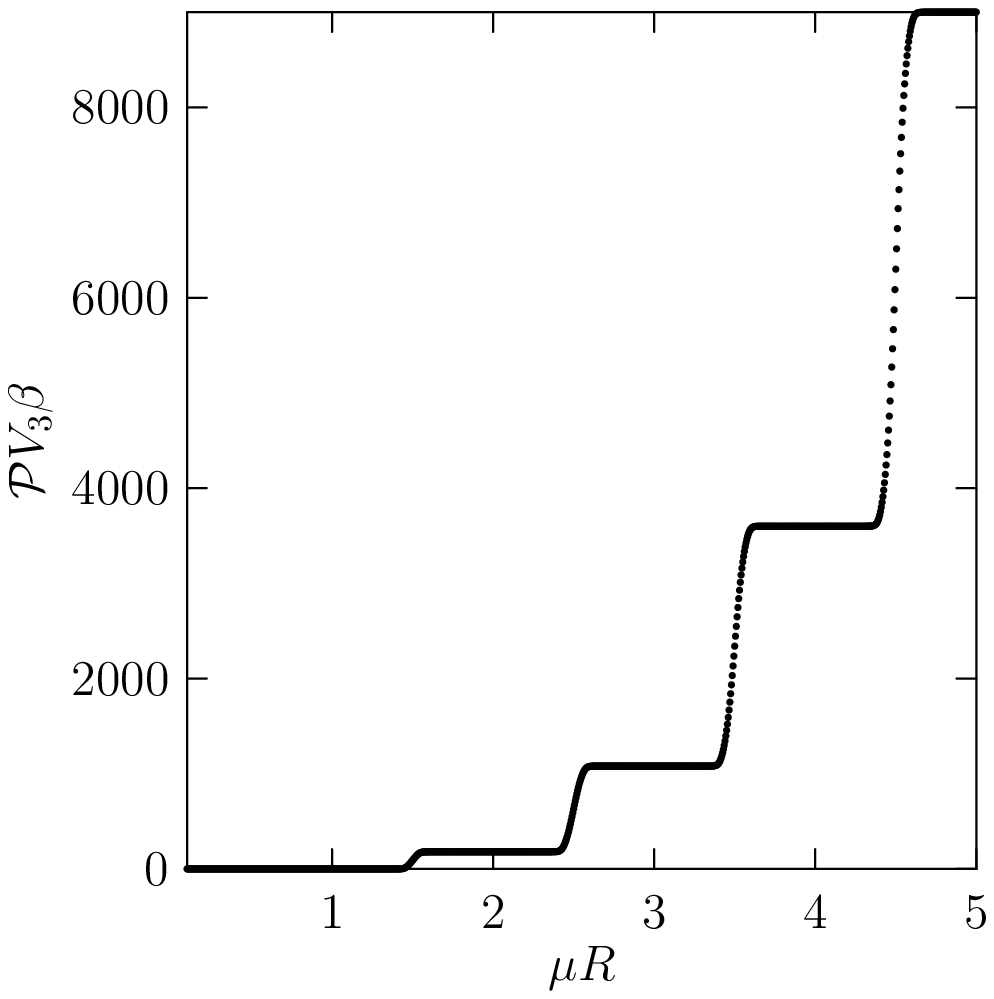}
    \end{center}
  \end{minipage}
  \hfill
  \begin{minipage}[t]{.49\textwidth}
    \begin{center}
\includegraphics[width=0.99\textwidth]{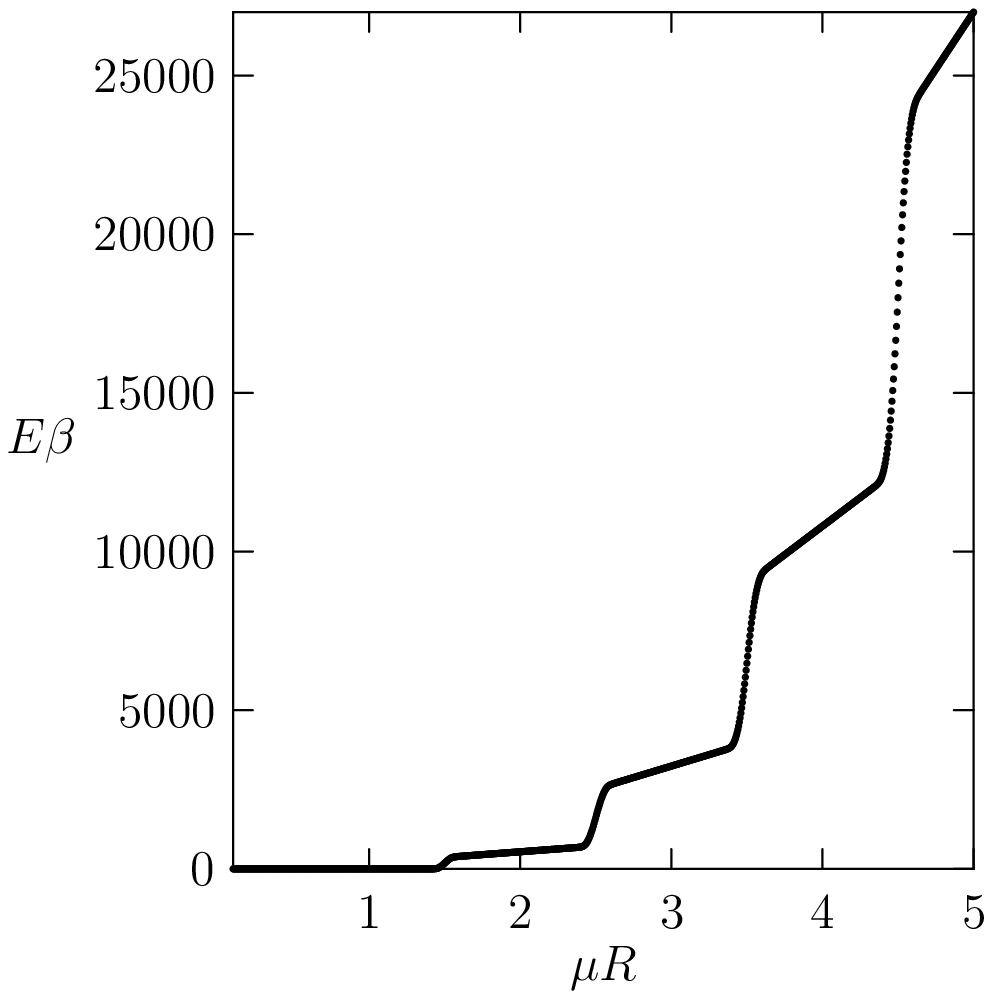}
    \end{center}
  \end{minipage}
  \hfill
\caption{(Left) Pressure and (Right) energy for $N = 3$, $N_f = 1$, $m = 0$, $\beta / R = 30$ (low $T$).}
\label{press_energy}
\end{figure}

The pressure indicates how the system responds to changes in the spatial volume. The pressure multiplied by the 4-volume is shown in Figure \ref{press_energy} (Left) for $N = 3$. In the $\beta \rightarrow \infty$, $m \rightarrow 0$ limit the expectation value of the pressure is given by
\EQ{
\begin{aligned}
{\cal P} \, & = \frac{1}{\beta} \left( \frac{\partial \log Z}{\partial V_3} \right)\\
&= \frac{-R}{3 \beta V_3 Z} \int \left[ {\mathrm d} \theta \right] e^{-S} \left( \frac{\partial S}{\partial R} \right)\\
&\xrightarrow[\beta \rightarrow \infty]{} \frac{N_f}{3 R V_3 Z} \int \left[ {\mathrm d} \theta \right] e^{-S} \sum_{\ell=1}^{\infty} \sum_{i=1}^{N} 2 \ell (\ell+1) (\ell+\tfrac12) \left[ \frac{e^{\beta \mu}}{e^{\beta \mu} + e^{-i \theta_i + \beta (\ell+\frac12) / R}} \right] ,
\end{aligned}
}
which indicates that each level has a pressure
\EQ{
{\cal P}_L = \frac{N N_f}{3 R V_3} \sum_{\ell=1}^L 2 \ell(\ell+1)(\ell+\tfrac12) .
}

\subsubsection{Energy $E$ for $m = 0$}

The energy can be evaluated from the pressure and fermion number above and is plotted in Figure \ref{press_energy} (Right). The energy is calculated as
\EQ{
\begin{aligned}
E &= - {\cal P} V_3 + \mu {\mathscr N} \\
&\xrightarrow[\beta \rightarrow \infty]{}\\
&\frac{N_f}{Z} \int \left[ {\mathrm d} \theta \right] e^{-S}
\sum_{\ell=1}^{\infty} \sum_{i=1}^{N} 2 \ell (\ell+1) \left( \mu -
  \tfrac{1}{3} 
(\ell+\tfrac12)/R \right) \left[ \frac{e^{\beta \mu}}{e^{\beta \mu} + e^{-i \theta_i + \beta (\ell+\frac12) / R}} \right].
\end{aligned}
}
This shows that the energy levels are not horizontal. The factor of $\mu$ in front of the fermion number in the first line causes the levels to rise linearly with $\mu$. The energy of each level is given by
\EQ{
E_L = N N_f \sum_{l=1}^{L} 2 \ell (\ell+1) \left( \mu - \tfrac{1}{3} (\ell+\tfrac12)/R \right) .
}

\subsubsection{Normalized ${\mathscr N}$, ${\cal P}$, $E$ for $m = 0$; the Stefan-Boltzmann limit}

\begin{figure}[t]
\center
\includegraphics[width=9cm]{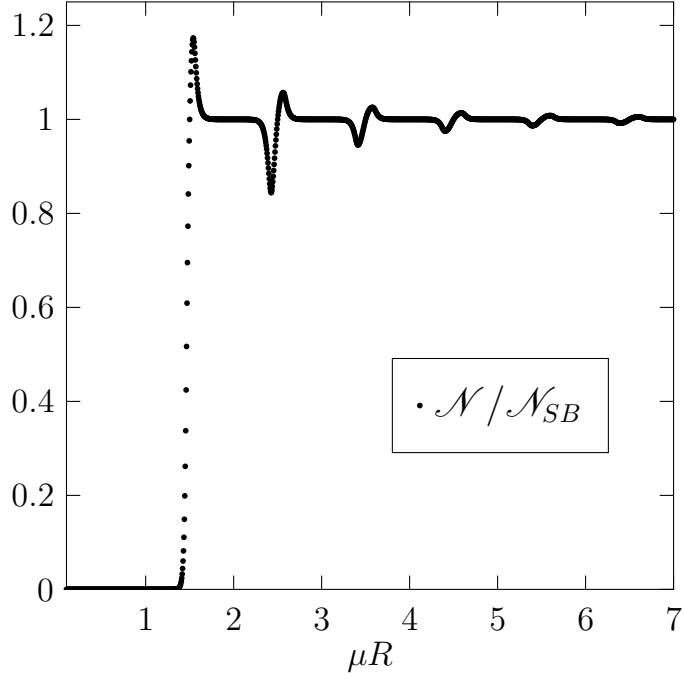}
\caption{Average fermion number normalized by its Stefan-Boltzmann value as a function of the chemical potential. $N = 3$, $m = 0$, $\beta / R = 30$ (low $T$).}
\label{norm_steps}
\end{figure}

The Stefan-Boltzmann limit is the zero interaction free fermion limit. On $S^1 \times S^3$ we obtain it from the one-loop result taking all the $\theta_i = 0$, corresponding to the deconfined phase, e.g., for the fermion number
\EQ{
{\mathscr N}_{SB} &\xrightarrow[\beta \rightarrow \infty]{} N N_f \sum_{\ell=1}^{\infty} 2 \ell(\ell+1) \left[ \frac{e^{\beta \mu}}{e^{\beta \mu} + e^{\beta (\ell+\tfrac12) / R}} \right] .
}
The normalized fermion number, ${\mathscr N} / {\mathscr N}_{SB}$, is shown in Figure \ref{norm_steps}. The behavior of ${\cal P} / {\cal P}_{SB}$ and $E/ E_{SB}$ as a function of $\mu$ is almost indistinguishable from ${\mathscr N} / {\mathscr N}_{SB}$. These results might at
first seem at odds with the results for $\PP_1$ and $\PP_{-1}$, in the
confined regions, since the $\theta_i$ are set to $0$ in ${\cal
  P}_{SB}$, ${\mathscr N}_{SB}$, and $E_{SB}$. The resolution is that
sufficiently far within the confined regions the observables are
independent of the $\theta_i$. When $\mu R$ is not close to $\ell + 
\tfrac{1}{2}$ for $l = 1, 2, ...$, then in the $\beta \rightarrow \infty$ limit the $\theta_i$ terms drop out of the exponentials in the Fermi-Dirac distributions,
\EQ{
i \theta_i + \frac{\beta}{R} \left(\ell+\tfrac{1}{2}\right) - \mu
\beta \xrightarrow[( \mu R \not\approx \ell+\frac12 )]{\beta
  \rightarrow \infty} \beta\left(\ell+\tfrac{1}{2}\right)R^{-1} - \mu
\beta. 
}

\subsubsection{Average Phase $\langle e^{i \phi} \rangle_{p q}$ for $m = 0$}

\begin{figure}[t]
\center
\includegraphics[width=9cm]{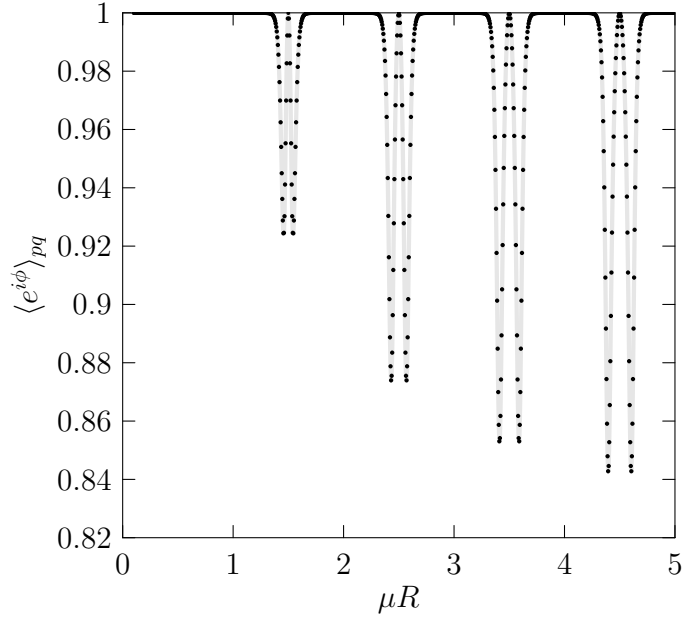}
\caption{Average phase as a function of chemical potential. $N = 3$, $m = 0$, $\beta / R = 30$ (low $T$). The average phase only differs from $1$ when $\PP_1 \ne \PP_{-1}^*$.}
\label{phase}
\end{figure}

The average phase is important in that it shows where the sign problem is severe. It is given by
\EQ{
\langle e^{i \phi} \rangle _{p q} \equiv \langle e^{-i {\rm Im} (S)} \rangle _{p q} = \frac{Z}{\int [{\mathrm d} \theta] e^{\RE(-S)}},
}
where the denominator is the ``phase quenched" (real action) partition 
function,  
\EQ{ 
Z_{pq} = \int [{\mathrm d} \theta] \left| e^{-S} \right| = \int [{\mathrm d} \theta] e^{\RE(-S)} .
}
Numerical results for the average phase are presented in Figure \ref{phase}. The results suggest that the sign problem increases in severity with the chemical potential, then levels off (large $\mu R$ results suggest it levels off near $0.83$). However, the average phase is only different from $1$ while $\PP_1$ and $\PP_{-1}^*$ differ. This explains why the average phase is always $1$ in the middle of a transition. In fact, We observe that $\langle e^{i \phi} \rangle_{p q}$ is smallest (largest) when $\left| \PP_1 - \PP_{-1}^* \right|$ is largest (smallest).


%
%
%
%
\subsection{Chiral condensate $\langle {\bar \psi} \psi \rangle$}

The chiral condensate is given by
\SP{
\langle {\bar \psi} \psi \rangle &= - \frac{1}{\beta V_3} \lim_{m \rightarrow 0} \left( \frac{\partial \log Z}{\partial m} \right)\\
&= \frac{1}{\beta V_3 Z} \lim_{m \rightarrow 0} \int \left[ {\mathrm d} \theta \right] e^{-S} \left( \frac{\partial S}{\partial m} \right)\\
&\xrightarrow[\beta \rightarrow \infty, m \rightarrow 0]{} \frac{N_f
  m}{\pi^2 R^2} \int \left[ {\mathrm d} \theta \right] e^{-S}
\sum_{\ell=1}^{\infty} \sum_{i=1}^{N} \frac{\ell
  (\ell+1)}{(\ell+\frac{1}{2})} \left[ \frac{e^{\beta \mu}}{e^{\beta
      \mu} + e^{-i \theta_i + \beta (\ell+\frac12) / R}} \right]\ .
}
In the massless limit $\langle {\bar \psi} \psi \rangle = 0$ as expected since it is a perturbative result, but it is interesting to note that in the light mass limit $\langle {\bar \psi} \psi \rangle$ is linear in $m$.


\subsection{Continuum results ($m R \rightarrow \infty$)}

Since all of our observables are functions of the dimensionless quantities $\beta / R$, $m R$, or $\mu R$, then we can obtain a ``continuum" limit by taking one of the following:

\begin{itemize}
\item $\beta / R$ small (high temperature perturbation theory),
\item $\mu R$ large (high density perturbation theory),
\item $m R$ large (heavy quarks).
\end{itemize}

\noindent We take $m R$ large to maintain validity in the region of
the confinement-deconfinement transitions. For sufficiently large mass
the fermion contribution to the action can be converted from a sum to
an integral using the Abel-Plana formula and then simplified by taking
$m R \rightarrow \infty$. This proceeds as 
\SP{
z_f (n \beta / R, m R) &= 2 \sum_{\ell=0}^{\infty} \ell ( \ell + 1 )
e^{- n \beta \sqrt{m^2+(\ell + \frac12)^2 R^{-2}}}\\ 
&= 2 \int_0^{\infty} {\mathrm d} y \, \left( y^2 - \tfrac{1}{4} \right)
e^{- \frac{n \beta}{R} \sqrt{y^2 + m^2 R^2}}\\ 
&~~~~~~~+4\int_{mR}^\infty dy\frac{y^2+\tfrac14}{e^{2\pi
    y}+1}\sin \left(n \beta \sqrt{y^2-m^2R^2}/R\right)\\
&\\
&\xrightarrow[m R \rightarrow \infty]{} 2 \int_0^{\infty} {\mathrm d}
y \, \left( y^2 - \tfrac{1}{4} \right) e^{- n \beta  \sqrt{y^2 + m^2
    R^2}/R} . 
\label{cont_formula}
}
The resulting integral formula removes all the periodic structure from the observables. It should be noted that this result is only valid in the vicinity of, $\mu R = m R$. Once the chemical potential becomes sufficiently large compared to the fermion mass then the periodic structure returns and the system behaves again as it did in the massless case.

\subsubsection{${\mathscr N}$ for $m \rightarrow \infty$}

\begin{figure}[t]
  \hfill
  \begin{minipage}[t]{.49\textwidth}
    \begin{center}
\includegraphics[width=0.99\textwidth]{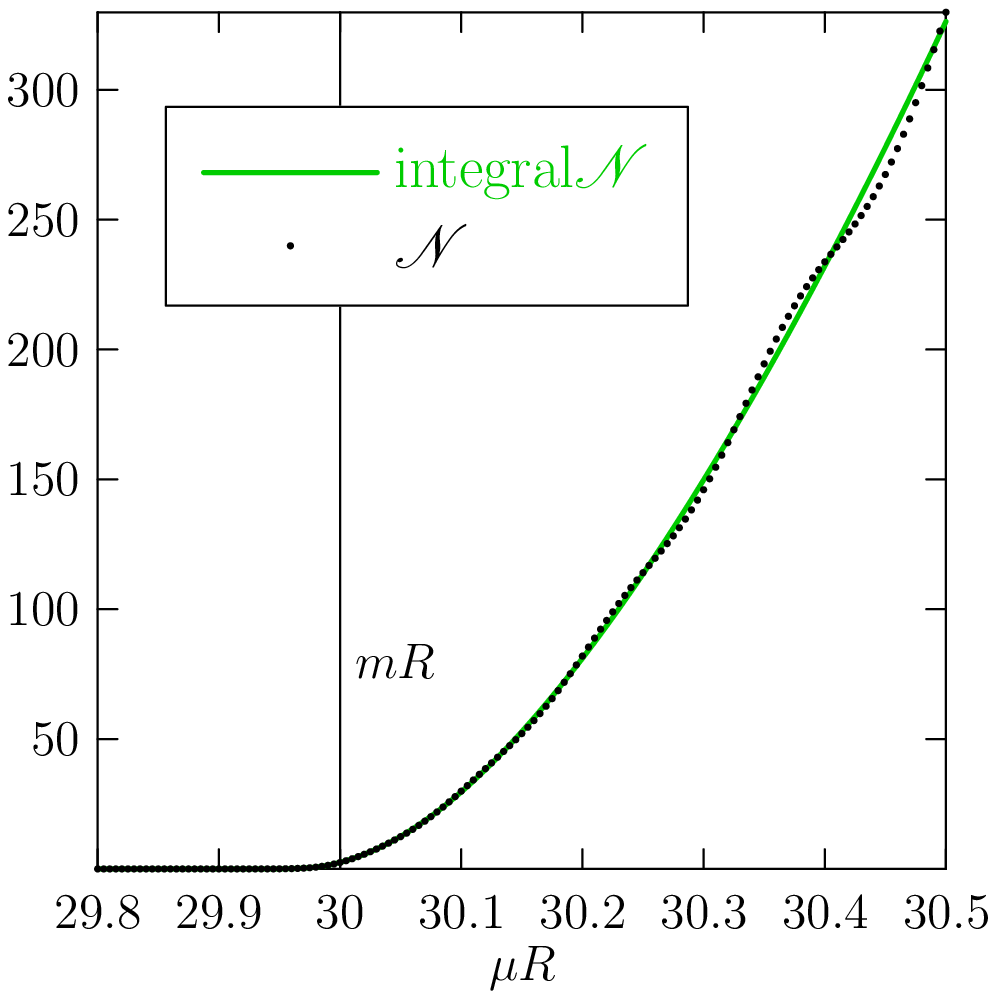}
\vspace{-2mm}
    \end{center}
  \end{minipage}
  \hfill
  \begin{minipage}[t]{.49\textwidth}
    \begin{center}
\includegraphics[width=0.99\textwidth]{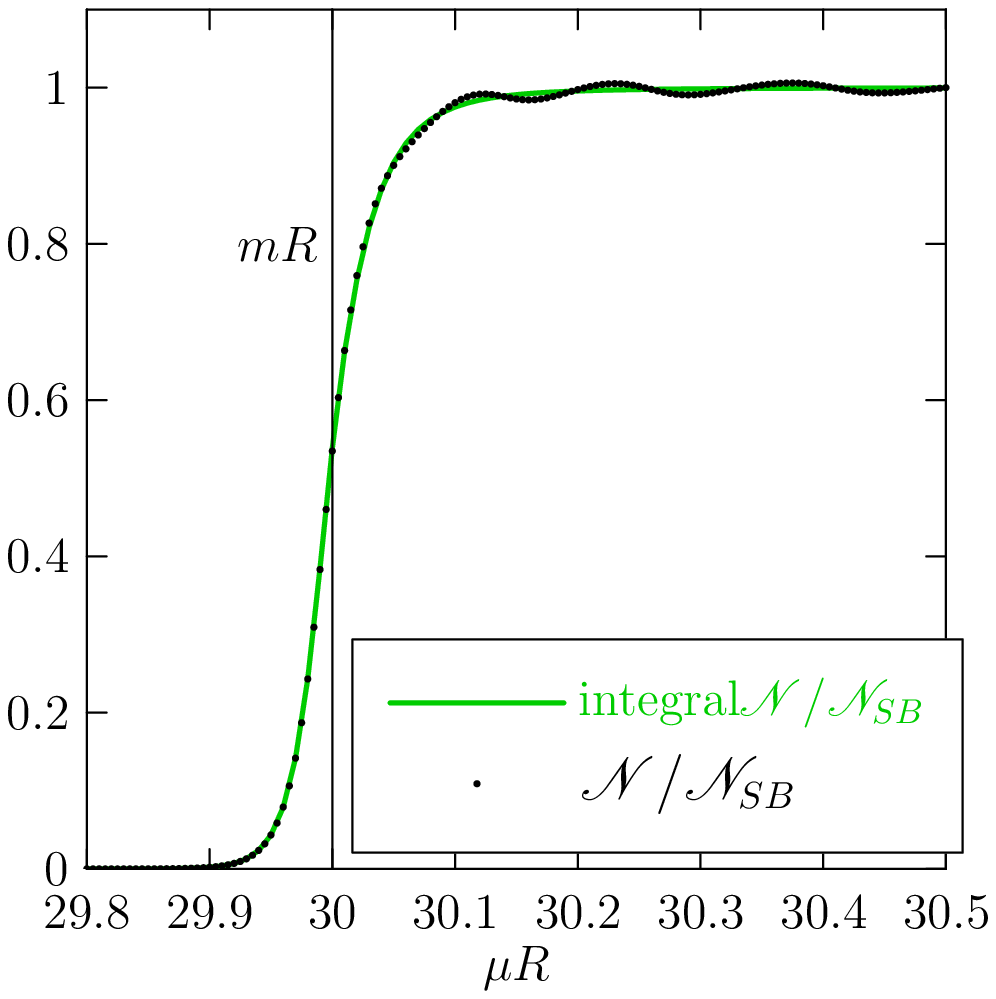}
\vspace{-2mm}
    \end{center}
  \end{minipage}
  \hfill
\caption{Fermion number: (Left) unnormalized, and (Right) normalized by its Stefan-Boltzmann value as a function of the chemical potential for large quark mass near onset at $\mu R = m R = 30$. $N = 3$, $N_f = 1$, $\beta / R = 30$ (low $T$). The dots are calculated using the full sum form of $z_f$. The curves are from the integral approximation.}
\label{fermnum_mass}
\end{figure}



For QCD with non-zero quark mass the expectation values ${\mathscr N}$, ${\cal P}$, and $E$ exhibit ``Silver Blaze" behavior: bulk observables are zero until onset \cite{Cohen:2003kd} which occurs when the chemical potential reaches the value of the lightest quark mass. Here onset occurs for $\mu = m$. The fermion number is given by
\EQ{
{\mathscr N} \xrightarrow[\beta \rightarrow \infty]{m R \rightarrow \infty} \frac{N_f}{Z} \int \left[ {\mathrm d} \theta \right] e^{-S} \int_{0}^{\infty} {\mathrm d}y \, 2 (y^2 - 1/4) \sum_{i=1}^{N} \left[ \frac{e^{\beta \mu}}{e^{\beta \mu} + e^{-i \theta_i + \beta  \sqrt{y^2 + m^2 R^2}R^{-1}}} \right] .
}
Each level $L$ has height
\EQ{
h_L = 2 N N_f \sum_{\ell=1}^{L} \ell(\ell+1) \hspace{1cm} \rightarrow \hspace{1cm} h_y = 2 N N_f \int {\mathrm d}y (y^2 - \tfrac14) ,
}
and level width
\EQ{
\begin{aligned}
(\Delta \mu R)_{\Delta l} &= \sqrt{\left(L + \tfrac{1}{2} + 1\right)^2 + m^2 R^2} - \sqrt{\left(L+\tfrac{1}{2}\right)^2 + m^2 R^2}\\
&\rightarrow \sqrt{(y+{\mathrm d}y)^2 + m^2 R^2} - \sqrt{y^2 + m^2 R^2}\\
&\rightarrow 0.
\end{aligned}
}
The Stefan-Boltzmann values shift accordingly as well, {\it e.g.\/}, for the fermion number
\EQ{
\begin{aligned}
{\mathscr N}_{SB} &\xrightarrow[\beta \rightarrow \infty]{} N N_f \sum_{\ell=1}^{\infty} 2 l (l+1) \left[ \frac{e^{\beta \mu}}{e^{\beta \mu} + e^{(\beta/R) \sqrt{(\ell+\frac12)^2 + m^2 R^2}}} \right]\\
&\xrightarrow[\beta \rightarrow \infty, m R \rightarrow \infty]{} N N_f \int_{0}^{\infty} {\mathrm d}y \, 2 (y^2 - \tfrac14) \left[ \frac{e^{\beta \mu}}{e^{\beta \mu} + e^{(\beta / R) \sqrt{y^2 + m^2 R^2}}} \right] .
\end{aligned}
}
The fermion number for large quark mass is plotted in Figure \ref{fermnum_mass}. The dots are calculated using the full sum form of $z_f$ (line $1$ of eq. (\ref{cont_formula})) and the green curves are from the integral approximation (line $4$). As expected ${\mathscr N}$ is close to $0$ until the onset transition at $\mu = m$. The smoothness of ${\mathscr N}$ as it increases from onset results from the decreased level width in the large $m R$ limit. It is only temporary though; as $\mu R$ is increased further from $m R$ the levels spread out and eventually the observables behave as they would for $m R = 0$. This is reflected in the oscillations that develop as $\mu R$ is increased when considering the full sum form of $z_f$. The lack of these oscillations in the integral form shows that this approximation breaks down for values of $\mu R$ away from the onset transition at $m R$. How long this breakdown takes depends on $m R$. The larger the $m R$, the further in $\mu R$ we can go before breakdown.

\subsubsection{Polyakov lines: $\PP_1$ and $\PP_{-1}$ and average phase $\langle e^{i \phi} \rangle_{p q}$ for $m \rightarrow \infty$}

\begin{figure}[t]
  \hfill
  \begin{minipage}[t]{.49\textwidth}
    \begin{center}
\includegraphics[width=0.99\textwidth]{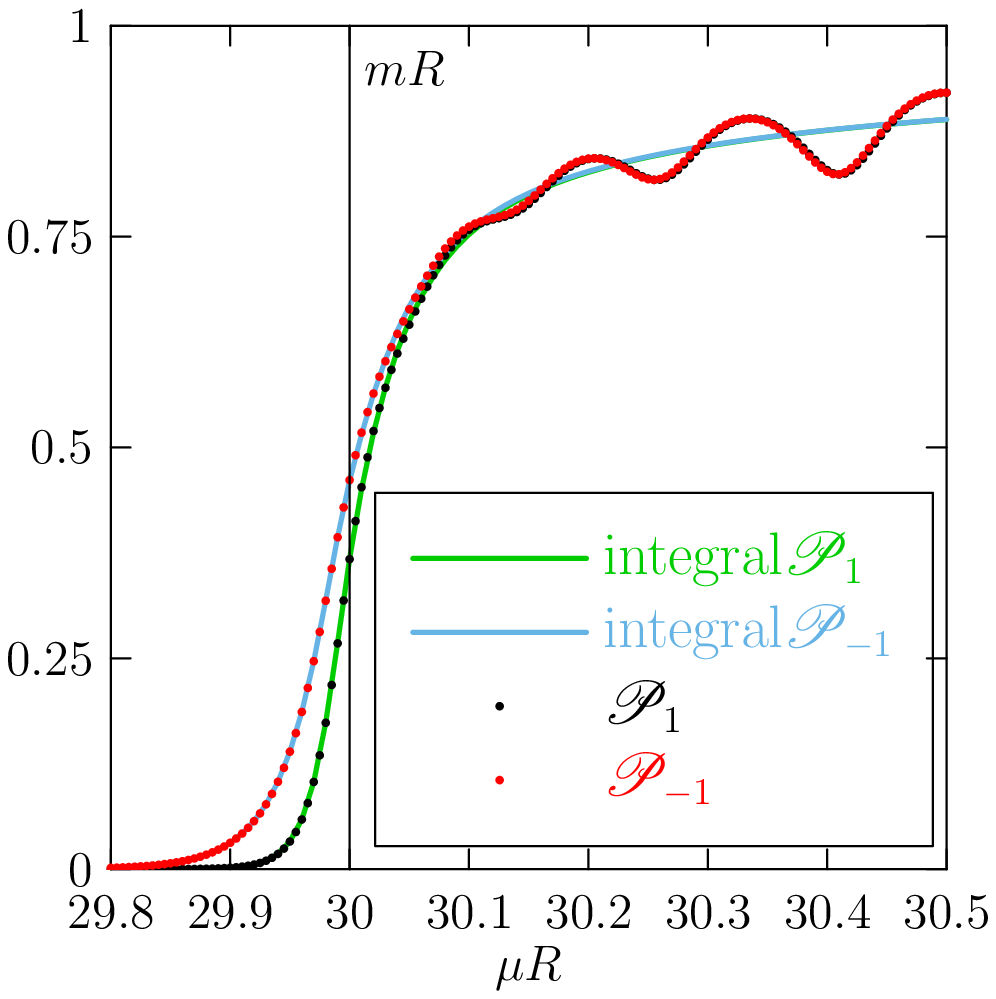}
    \end{center}
  \end{minipage}
  \hfill
  \begin{minipage}[t]{.49\textwidth}
    \begin{center}
\includegraphics[width=0.99\textwidth]{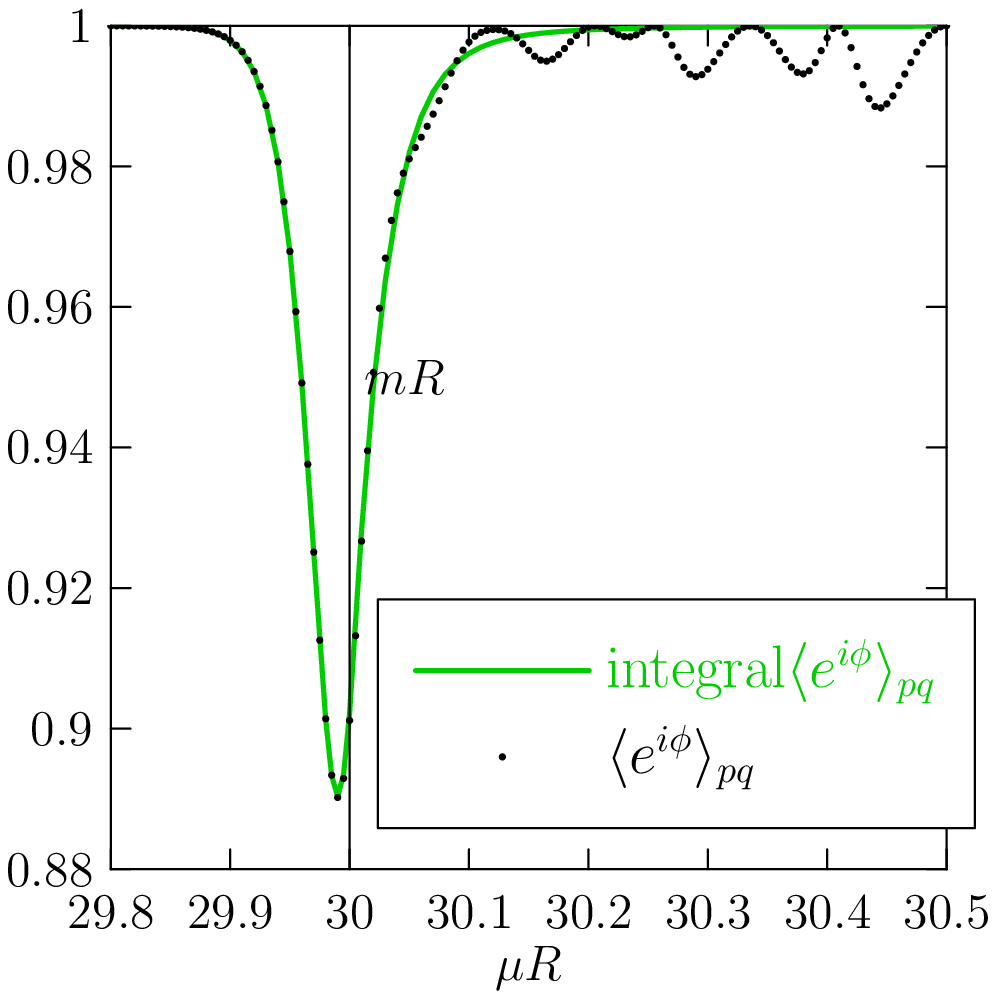}
    \end{center}
  \end{minipage}
  \hfill
\caption{(Left) $\PP_1$ and $\PP_{-1}$ as a function of chemical potential for large quark mass near onset at $\mu R = m R = 30$. $N = 3$, $N_f = 1$ $\beta / R = 30$ (low $T$). The dots are calculated using the full sum form of $z_f$. The curves are from the integral approximation. (Right) The average phase is calculated under the same conditions.}
\label{poly_phase_trans_mass}
\end{figure}

As in the $m R = 0$ discrete case, in the large $m R$ limit the behavior of $\PP_{-1}$ precedes that of $\PP_{1}$ as a function of $\mu R$. The transition in $\mu R$ occurs around onset at $m R$ and appears sharper for larger $m R$. Results for $m R = 30$ near the transition are presented in Figure \ref{poly_phase_trans_mass} (Left), which shows $\PP_{1}$ and $\PP_{-1}$ using the sum form of $z_f$ (black dots) and the integral approximation (green and blue curves). It is clear again that the integral approximation breaks down after the transition.

Figure \ref{poly_phase_trans_mass} (Right) shows the average phase near the transition. After the transition, and before $\mu R$ is sufficiently large that oscillations return full force, there is a brief respite from the sign problem in that the average phase is close to $1$. The small spike in $\langle e^{i \phi} \rangle_{p q}$ at $\mu = m$ corresponds to the value of $\mu R$ where $\PP_1$ and $\PP_{-1}^*$ differ maximally and serves as a good indicator of the location of the transition. Eventually, regardless of how large we make $m R$, the oscillations always return by taking $\mu R$ sufficiently bigger.



\section{The Large $N$ Theory at Low $T$}

In this section we turn to an analysis of the large-$N$ theory,
specifically in the low-temperature limit $T\ll R^{-1}$. We showed in
\eqref{mmod} that in this limit the system reduced to a unitary matrix
model with a potential determined by the quarks.\footnote{A similar 
matrix model but defined in the canonical ensemble and truncated so as to
include only the $n=1$ term was considered in 
\cite{Azakov:1986pn}. We work directly in the grand canonical ensemble
and it is important in our analysis that we do not truncate the sum
over $n$ in the fermionic sector.}

Without the chemical potential, the potential $V(P)$ in \eqref{mmod} 
would vanish in the
low temperature limit. In this case, the Vandermonde piece dominates
leading to a repulsion of the eigenvalues and so the integral
is dominated by a uniform distribution of the angles $\theta_i$ around
the circle. Consequently $\langle \Tr
P^n\rangle=0$ and the theory is in the
confining phase. However, if as $T\to0$ we simultaneously 
tune $\mu\to\varepsilon$ keeping $\beta(\mu-\varepsilon)$
fixed, where $\varepsilon$ is one of the energy levels
$\varepsilon_\ell$ of the fermion
system then the fermion term can compete with the van de Monde piece
and a phase transition can occur. In fact this is a version of 
the Gross-Witten
transition \cite{Gross:1980he} (see also 
\cite{Wadia:1979vk,Wadia:1980cp}). The Gross-Witten transition is typically a 
third order phase transition that
occurs in large $N$ gauge theories in finite volume and involves a
transition from a configuration where the density of eigenvalues of
the Polyakov line $\{e^{i\theta_i}\}$ lies on the unit circle in the
complex $z=e^{i\theta}$ plane to a configuration with a gap. We
loosely refer to this as a confinement/deconfinement transition
because the confined phase defined by a distribution that lies on a closed contour is smoothly connected to the phase where the eigenvalues are
uniformly distributed around the unit circle and $\langle
\Tr P^n\rangle=0$, for $n\neq0$, indicative that it costs an infinite amount of
energy to propagate colour charges. 
Correspondingly the deconfined phase defined by a distribution that has a gap is
smoothly connected to the configurations where all the eigenvalues
$e^{i\theta_i}=1$, {\it i.e.\/}~$\langle \Tr P^n\rangle=N$ where colour
charges are free to propagate indicative of a plasma phase.
We will find that, as suggested by the $N=3$ results, 
increasing the chemical potential induces a series of
such transitions,
the novel aspect being that the eigenvalues no longer lie on the unit
circle due to the complex action.

Without-loss-of-generality we will choose $\mu$ to be
positive so that only the contribution from the quarks survive and the
quark potential takes the form
\EQ{
V(P)=-\sum_\ell\sigma_\ell\log\Big(1+e^{
\beta(\mu-\varepsilon_\ell)}P\Big)\ .
\label{act3}
}
Notice that the potential is not hermitian.
This is the usual ``sign
problem'' in the presence of a chemical potential. In the context of
the matrix model, it means that the saddle point that dominates in the
large $N$ limit will lie at complex angles $\theta_i$. In this respect
the matrix model is of the ``holomorphic'' type that appears in the
Dijkgraaf-Vafa approach to supersymmetric gauge theories and in
particular \cite{Dijkgraaf:2002vw} which considered the unitary version.

\subsection{Single level model}

For the moment let us focus on what can happen with a single level
$\varepsilon\equiv\varepsilon_1$ and $\sigma\equiv\sigma_1$. Defining 
an effective fugacity $\xi=e^{\beta(\mu-\varepsilon)}$ the effective
action on the angles, including the Vandermonde piece, is
\EQ{
S(\theta_i)=-\frac12\sum_{i,j=1}^{N}\log\sin^2\Big(\frac
{\theta_i-\theta_j}2\Big)+N\sum_{i=1}^{N}V(\theta_i)\ ,~~~
V(\theta)=i\kN\theta-\sigma\log\left(1+\xi e^{i\theta}
\right)\ ,
\label{act4}
}
where we have added the Lagrange multiplier $\kN$ to enforce the
$\det P=1$ constraint, {\it i.e.\/}~$\sum_i\theta_i=0$.

In the large $N$ limit, the integral over the angles
is dominated by a saddle point obtained by solving the
equation-of-motion that follows from \eqref{act4}
\EQ{
i\kN-\frac{i\sigma\xi e^{i\theta_i}}{1+\xi e^{i\theta_i}}
=\frac1N\sum_{j(\neq i)}\cot\left(\frac{\theta_i-\theta_j}2\right)\ .
\label{eom6}
}
As we remarked above, our system is a unitary matrix
model with a non-standard kind of potential term $V(\theta)$ coming from the
fermions. What is novel in the present context is that the
potential is not real and as a consequence the saddle-point
configuration will lie out in the complex plane. In fact if we define
$z_i=e^{i\theta_i}$ then in the presence of the non-real potential the
$z_i$ will move off the unit circle in the $z$-plane. 
As a consequence if we define the observables
\EQ{
\PP_n=\langle\Tr P^n\rangle=\frac1{N}\sum_{i=1}^{N}e^{in\theta_i}\ ,
}
we will find that $\PP_{-n}\neq\PP_n^*$ on the saddle point solution.

Before we plunge in and solve the matrix model, it is useful to note
that when $\xi$ is either very small or large, the potential vanishes
(in the latter case following from using the constraint
$\sum_i\theta_i=0$) 
and so we expect the $\{z_i\}$ to be uniformly distributed around the unit
circle. Taking the equation-of-motion in terms of the $z_i$ variables
\EQ{
\kN-\frac{\sigma\xi z_i}{1+\xi z_i}=\frac1N\sum_{j(\neq
  i)}\frac{z_i+z_j}{z_i-z_j}\ ,
\label{eom1}
}
it then follows that as $\xi\to0$ the Lagrange multiplier
$\kN=0$ while as $\xi\to\infty$ the Lagrange multiplier 
$\kN=\sigma$. The Lagrange multiplier
$\kN$ has an important interpretation following from \eqref{eom1}:
\EQ{
\kN=\frac1N\sum_i\frac{\sigma\xi z_i}{1+\xi z_i}=
\frac T{N^2}\frac{\partial\log Z}{\partial\mu}
\label{rer}
}
and so $\kN$ is the effective fermion number, $\frac{{\mathscr N}}{N^2}$. So
as $\mu$ varies from $\mu\ll\varepsilon$ to $\mu\gg\varepsilon$ the
picture is that the energy 
level becomes occupied and the effective fermion number 
jumps by the factor
$\sigma$. The question
before us is to establish how this transition occurs. 

{\sl The small $\xi$ confined phase}

We make the hypothesis that as $\xi$ increases from 0 that the
eigenvalues are continuously distributed around 
a closed contour ${\cal C}$ in the $z$-plane at
least up to some finite value of $\xi$. In the large $N$ limit, we can
describe the distribution of eigenvalues 
by an analytic function $\varrho(z)$ for which $\varrho(z)dz$
along ${\cal C}$ is real and positive. It is useful to
think of $\varrho(z)$ in terms of a conformal map of the
cylinder $z(s)$,  $-\pi\leq s\leq\pi$, $s \in {\field R}$, for which the eigenvalues are
uniformly distributed along the real axis between $-\pi$ and $\pi$, to 
the $z$-plane:
\EQ{
\frac1N\sum_{i=1}^{N}\longrightarrow \int_{-\pi}^\pi \frac{ds}{2\pi}
=\oint_{{\cal C}}\frac{dz}{2\pi i}\varrho(z)\ ,
}
where the contour is obtained from the inverse map $z(s)$ 
obtained by solving the differential equation 
\EQ{
i\frac{ds}{dz}=\varrho(z)\ ,
\label{dffe}
}
subject to the initial condition that when $\xi=0$ we have $z=e^{is}$.
For consistency we must also have $z(-\pi)=z(\pi)$ so that ${\cal C}$
is closed. The normalization condition is
\EQ{
\oint_{{\cal C}}\frac{dz}{2\pi i}\varrho(z)=1\ ,
\label{norm}
}
where the left-hand side can be evaluated using Cauchy's theorem.
The $SU(N)$ condition, $\sum_i \theta_i = 0$, translates into the constraint
\EQ{
\int_{{\cal C}}\frac{dz}{2\pi i}\varrho(z)\log z=0\ ,
\label{con}
}
where the branch cut of the logarithm is taken through the contour
${\cal C}$ at the point $z(\pm\pi)$.

In the large-$N$ limit
the saddle-point equation \eqref{eom1} becomes for $z\in{\cal C}$
\EQ{
zV'(z)={\mathfrak P}\oint_{{\cal
    C}}\frac{dz'}{2\pi i}\,\varrho(z')
\frac{z+z'}{z-z'}\ ,~~~~~ zV'(z)=
\kN-\frac{\sigma\xi z}{1+\xi z}\ .
\label{eom2}
}
The ${\mathfrak P}$ here indicates a principal value which is the
required prescription given that
$z$ lies on integration contour ${\cal C}$. In practical terms this means
that the right-hand side is the average of two terms with $z$
infinitesimally just inside and just outside the integration
contour. The right-hand side can then be evaluated by Cauchy's theorem
yielding an expression of the form
$-z\varrho(z)+\cdots$ where
the ellipsis represent other terms that arise when $\varrho(z)$ has
other poles inside ${\cal C}$. It is clear that $\varrho(z)$ can only
have poles at $z=-\tfrac1\xi$ and $0$. The condition \eqref{norm}
implies that the pole at $0$ must be inside ${\cal C}$ and have unit residue.
The only question is whether
the pole at $z=-\tfrac1\xi$ is inside ${\cal C}$ or not. 
In the small $\xi$ region, it will be outside---indeed this will
define the small $\xi$ region. Consequently, from \eqref{rer}
\EQ{
\kN=\oint_{{\cal C}}\frac{dz}{2\pi i}\frac{\sigma\xi z\varrho(z)}{
1+\xi z}=0\ .
}
The form of $\varrho(z)$ is then 
completely fixed by \eqref{eom2} to be
\EQ{
\varrho(z)=\frac1z+\frac{\sigma\xi}{1+\xi z}\ .
}
One can check that the constraint \eqref{con} is satisfied.
Notice that as $\xi\to0$, $\varrho(z)\to\tfrac1z$, which is the uniform
distribution around the unit circle.
The contour ${\cal C}$ follows by solving \eqref{dffe} for $s$ to give
\EQ{
e^{is}=z(1+\xi z)^\sigma
\label{sas1}
}
and then inverting to give $z(s)$.
Notice that the integration constant is fixed by requiring that as
$\xi\to0$ we have $z=e^{is}$ (the uniform distribution of eigenvalues).
In this small $\xi$ confining phase 
the effective fermion number vanishes $\kN=0$ and the Polyakov line
expectation values are
\EQ{
\PP_1=\int_{{\cal C}}\frac{dz}{2\pi i}\varrho(z)z=0\ ,~~~~~
\PP_{-1}=\int_{{\cal C}}\frac{dz}{2\pi i}\varrho(z)\frac1z=\sigma\xi\ .
\label{dsd1}
}
As advertised, as a symptom of having a complex action, $\PP_{-1}\neq\PP_1^*$.

The question is what happens to this phase as $\xi$ increases. The
issue is that the pole of $\varrho(z)$ at $z=-\tfrac1\xi$ must lie outside
${\cal C}$, however, this cannot be maintained when
$\xi$ becomes large enough. To see what happens notice that
$\varrho(z)$ vanishes at $z=-\tfrac1{\xi(1+\sigma)}$, a point that lies
outside ${\cal C}$ for small $\xi$. As $\xi$ increases to the value
\EQ{
\xi=\xi_1=\frac{\sigma^{\sigma}}{(1+\sigma)^{1+\sigma}}
}
$\varrho(z)$ vanishes precisely on ${\cal C}$ at the point $z(\pm\pi)$
on the negative real axis. At this point ${\cal C}$ develops a kink
and beyond this value of $\xi$ the inverse solution
$z(s)$ to \eqref{sas1} ceases to be closed $z(-\pi)\neq z(\pi)$. This
signals that a phase transition will occur to a configuration where the contour
opens into an arc, just as in the matrix model solved by
Gross and Witten \cite{Gross:1980he}. The line of the phase transitions in the
$(\mu,T)$ plane corresponds to the straight line
\EQ{
\mu=\varepsilon-T\big[(1+\sigma)\log(1+\sigma)-\sigma\log\sigma\big]\
,
}
valid in the low temperature limit.

{\sl The large $\xi$ confined phase}

We can solve for the large $\xi$ phase in a similar way. In this case
the pole of $\varrho(z)$ at $z=-\tfrac1\xi$ is now inside ${\cal C}$.
The solution turns out to be
\EQ{
\varrho(z)=\frac{1+\sigma+\xi z}{z(1+\xi z)}\ ,
}
which satisfies \eqref{eom2} and \eqref{norm} with
$\kN=\sigma$. In addition, one can check that the constraint \eqref{con} is
satisfied. 
Notice that as $\xi\to\infty$, we have $\varrho(z)\to\tfrac1z$, as required.
The contour ${\cal C}$ follows by solving \eqref{dffe} to get
\EQ{
e^{is}=\frac{z^{1+\sigma}\xi^\sigma}{(1+\xi z)^\sigma}\ .
\label{sas2}
}
In this phase the effective fermion number $\kN=\sigma$---the level is now
occupied---and the Polyakov line expectation values are
\EQ{
\PP_1=\frac\sigma\xi\ ,~~~~~~\PP_{-1}=0\ .
}
So comparing with \eqref{dsd1}, the behaviour of $\PP_{\pm1}$ swaps over
along with the replacement $\xi\to\xi^{-1}$.

As in the small $\xi$ phase, the large $\xi$ 
phase persists until the zero of
$\varrho(z)$, at $z=-\tfrac{1+\sigma}\xi$ which lies inside ${\cal C}$ for
large enough $\xi$, just touches ${\cal C}$. This occurs when 
\EQ{
\xi=\xi_2=\frac{(1+\sigma)^{1+\sigma}}{\sigma^{\sigma}}\ .
}
For smaller values of $\xi$, the contour ${\cal C}$ is not closed and
the phase does not exist. Notice that the points of transition
$\xi=\xi_1$ and $\xi=\xi_2$ satisfy $\xi_1\xi_2=1$. In the $(\mu,T)$
plane the boundary lies along the straight line
\EQ{
\mu=\varepsilon+T\big[(1+\sigma)\log(1+\sigma)-\sigma\log\sigma\big]\
,
}
valid in the low temperature limit.

{\sl The deconfined phase}

\begin{figure}[t]
\centerline{\includegraphics[width=8cm]{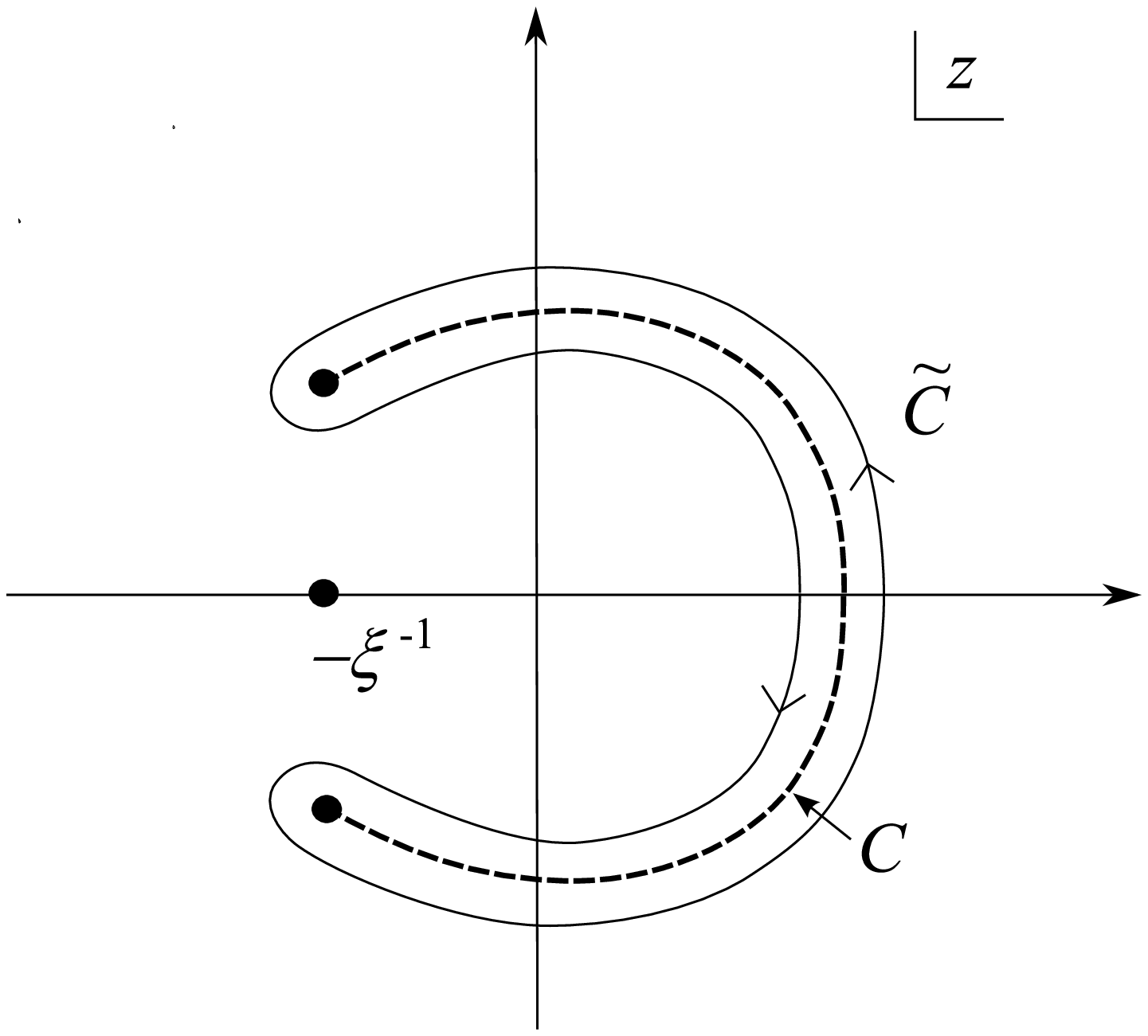}}
\caption{The resolvant $\omega(z)$ is naturally defined on the cut
  complex $z$-plane with the eigenvalues taking support along the cut
  ${\cal C}$.}
\label{f1}
\end{figure}

In the region $\xi_1\leq\xi\leq\xi_2$, experience with the
Gross-Witten matrix model suggests that the
eigenvalues lie on an open contour ${\cal C}$. In this case, the model
can be solved by the standard resolvant/spectral curve method. The
resolvant is defined as the function
\EQ{
\omega(z)=-\frac1N\sum_j \frac{z+z_j}{z-z_j}\ .
}
By hypothesis, in the large $N$ limit the eigenvalues distribute
themselves on an open contour ${\cal C}$ and $\omega(z)$ is
a function which is analytic everywhere in the $z$-plane except along a
square-root branch cut running along ${\cal C}$ with branch
points, $\tilde z$ and  $\tilde z^*$, lying at the endpoints, as illustrated in Figure \ref{f1}. The
eigenvalues are then distributed along the cut between the branch points, such that
\EQ{
\omega(z)=-\int_{{\cal C}}\frac{dz'}{2\pi i}\varrho(z')
\frac{z+z'}{z-z'}\ .
\label{res}
}
it follows from this representation that as $|z|\to0$ and $\infty$
\EQ{
\lim_{|z|\to0}\omega(z)=1\ ,~~~~\lim_{|z|\to\infty}\omega(z)
=-1\ . 
\label{lim}
}

From the Plemelj formulae \cite{musk:2008jrmr}, the equation-of-motion \eqref{eom2} is simply the condition 
\EQ{
zV'(z)=-\frac 12\big[\omega(z+\epsilon)+\omega(z-\epsilon)\big]
\ ,~~~~z\in{\cal C}\ ,
\label{eom4}
}
where $\epsilon$ is an infinitesimal such that $z\pm\epsilon$ lies on
either side of the cut.\footnote{This is a definition of the principal
  value in \eqref{eom2}.} It follows from \eqref{res} 
and the Plemelj formulae that 
the density of eigenvalues is obtained as the discontinuity of
$\omega(z)$ across the cut ${\cal C}$:
\EQ{
z\varrho(z)=\frac 1{2}\big[\omega(z+\epsilon)-\omega(z-\epsilon)\big]\
,~~~~z\in{\cal C}\ .
}
In particular, this means that an average of the form
\EQ{
\int_{{\cal C}}\frac{dz}{2\pi i}\varrho(z)F(z)=
\oint_{\tilde{\cal C}}\frac{dz}{4\pi i z}\omega(z)F(z)\ ,
}
can be used where $\tilde{\cal C}$ is a contour which encircles ${\cal C}$ as
illustrated in Figure \ref{f1}.
This can then be evaluated by pulling the closed contour $\tilde{\cal
  C}$ off the cut to pick up the residues in the $z$-plane.

The resolvant can be written down based on the following conditions:
it is an analytic function 
in the $z$-plane apart from a single square-root
branch cut, it satisfies the conditions \eqref{lim}, 
and $\varrho(\tilde z) = \varrho(\tilde z^*) = 0$. The solution
to the equation-of-motion \eqref{eom4} is of the form
\EQ{
\omega(z)=-zV'(z)+f(z)\sqrt{(z-\tilde z)(z-\tilde z^*)}\
,~~~~z\varrho(z)=f(z)\sqrt{(z-\tilde z)(z-\tilde z^*)}\ .
}
The endpoints $\tilde z$, $\tilde z^*$ and the function $f(z)$ are then determined by
imposing \eqref{lim} and requiring that $\omega(z)$ is regular at
$z=-\tfrac1\xi$. These conditions uniquely determine
\EQ{
f(z)=\frac{\sigma}{(1+\xi z)\big|\frac1\xi+\tilde z\big|}\ ,
}
and the endpoints of ${\cal C}$ are given by
\EQ{
\tilde z= \frac{-1}{\xi \left( 1 + \sigma - {\cal N} \right)^2} \left[ {\cal N}^2 + 1 + \sigma - {\cal N} \sigma +2 i \sqrt{ {\cal N} \left( \sigma - {\cal N} \right) \left( 1 + \sigma \right)} \right] .
} 
What remains is to fix $\kN$ by imposing the final
$SU(N)$ condition \eqref{con} which becomes
\EQ{
\oint_{\tilde{\cal C}}\frac{dz}{4\pi i z}\,\omega(z)\log z=0\ .
}
Pulling the contour off the cut, deforming it around the poles at $z = 0$ and
$z=-\tfrac1\xi$ collecting their residues, and adding the discontinuity from the branch cut of  
$\log z$ gives\footnote{For another example see \cite{Arsiwalla:2005jb}.}
\SP{
\oint_{\tilde{\cal C}}\frac{dz}{2\pi i z}\,\omega(z)\log z= 
\lim_{\epsilon\to0\atop\eta\to\infty}&\left[ \oint_{0}
\frac{dz}{2\pi i z}\,\omega(z)\log z-\oint_{\infty}
\frac{dz}{2\pi i z}\,\omega(z)\log z\right.\\
&\left.+\oint_{-1/\xi}\frac{dz}{2\pi i z}\,\omega(z)\log z
 + \int_{-\eta}^{-\epsilon}\frac{dz}z\,\omega(z) \right] = 0.
}
where $\epsilon$ and $\eta$ are cut-offs and the integrals
around $z=0$ and $\infty$ are defined on the contours
$\epsilon e^{i\theta}$ and $\eta e^{i\theta}$, $0\leq\theta<2\pi$,
respectively. The limit is well defined since the factors of 
$\log\epsilon$ and $\log\eta$ separately cancel. What remains is the condition
\EQ{
\xi=\frac{(\sigma-\kN)^{\sigma-\kN}(1+\kN)^{1+\kN}}
{\kN^\kN(1+\sigma-\kN)^{1+\sigma-\kN}}\ .
\label{mxx}
}
This equation determines $\kN$ as a function of $\xi$. An important
check on the solution is how the deconfined phase interfaces with the 
small and large $\xi$ closed phases. When 
$\kN=0$ we have $\xi=\xi_1$ and at this point $\IM \tilde z=0$ and so
the two branch points come together and one can readily verify that
the density $\varrho(z)$ matches that of the small $\xi$ phase at the
edge of its existence at
$\xi=\xi_1$. The same happens at $\kN=\sigma$ when $\xi=\xi_2$ where the
gapped phase merges continuously with the large $\xi$ closed phase.

\begin{figure}[ht] 
\centerline{\includegraphics[width=3.5in]{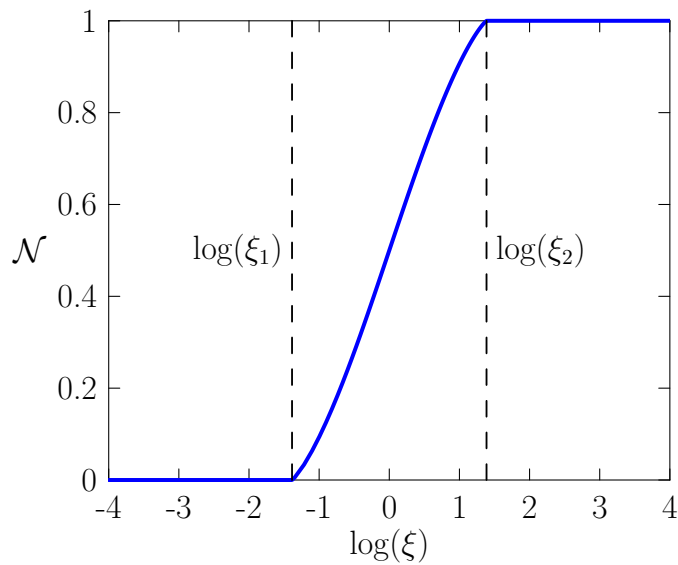}}
\caption{\small The effective fermion number across the pair of
  Gross-Witten transitions 
  from the small $\xi$ confined phase through the 
deconfined phase to the
  large $\xi$ confined phase.}\label{fig3}
\end{figure}

From \eqref{mxx} it follows that across the transitions at $\xi=\xi_1$
and $\xi=\xi_2$,
$\kN$ and its first derivative $\tfrac{\partial\kN}{\partial\mu}$ are
continuous, however higher derivatives are discontinuous. Since $\kN$ is
the effective fermion number, the first derivative of the grand potential, 
it follows that the transitions are third order just as in the original
Gross-Witten model \cite{Gross:1980he}. The behaviour of the 
effective fermion number as $\mu$ is increased past $\varepsilon$ is
shown in Figure \ref{fig3}.

\begin{figure}[ht] 
\centerline{\includegraphics[width=3.5in]{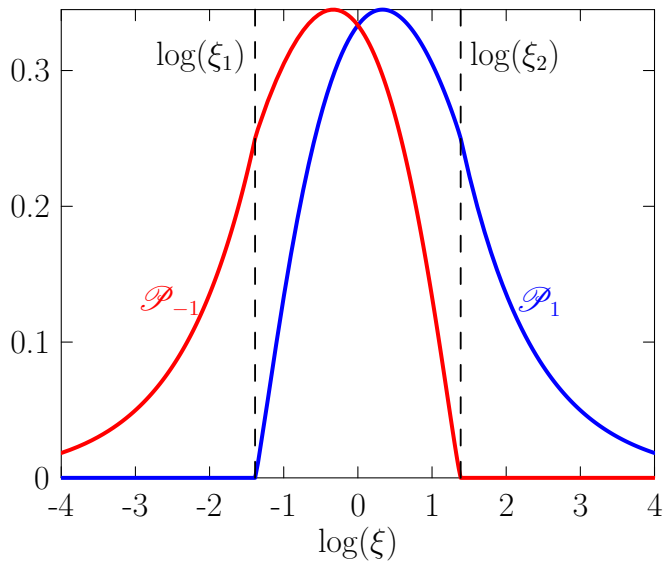}}
\caption{\small The Polyakov lines $\PP_1$ (in blue) and $\PP_{-1}$ (in
  red) from the small to large $\xi$ phases via the deconfined phase. 
The transitions from the confined/deconfined phases occur when either $\PP_1$
or $\PP_{-1}$ vanish.}\label{fig2}
\end{figure}

The expectation values of the Polyakov line in the deconfined 
phase can be determined from an expansion of the resolvant for $z \rightarrow 0$ and $z \rightarrow \infty$\cite{Semenoff:2004bs},
\EQ{
\omega(z) = -1 - 2 \sum_{n=1}^{\infty} \frac{1}{z^n} \PP_n ,
}
\EQ{
\omega(z) = 1 + 2 \sum_{n=1}^{\infty} z^n \PP_{-n} .
}
where $n$ is the number of windings. For a single winding, the Polyakov lines are
\EQ{
\PP_1=\frac{\kN}{\sigma+1-\kN}\frac1\xi\ ,~~~~~~~
\PP_{-1}=\frac{\sigma-\kN}{1+\kN}\xi
\ ,
}
where $\kN=\kN(\xi)$ via the inversion of \eqref{mxx}
and $\xi=e^{\beta(\mu-\varepsilon)}$. The behaviour of $\PP_{\pm1}$ is
shown in Figure \ref{fig2}. Notice how $\PP_{\pm1}$ vanish in the
small and large $\xi$ confined phases, respectively.

\begin{figure}[ht] 
\centerline{\includegraphics[width=5.9in]{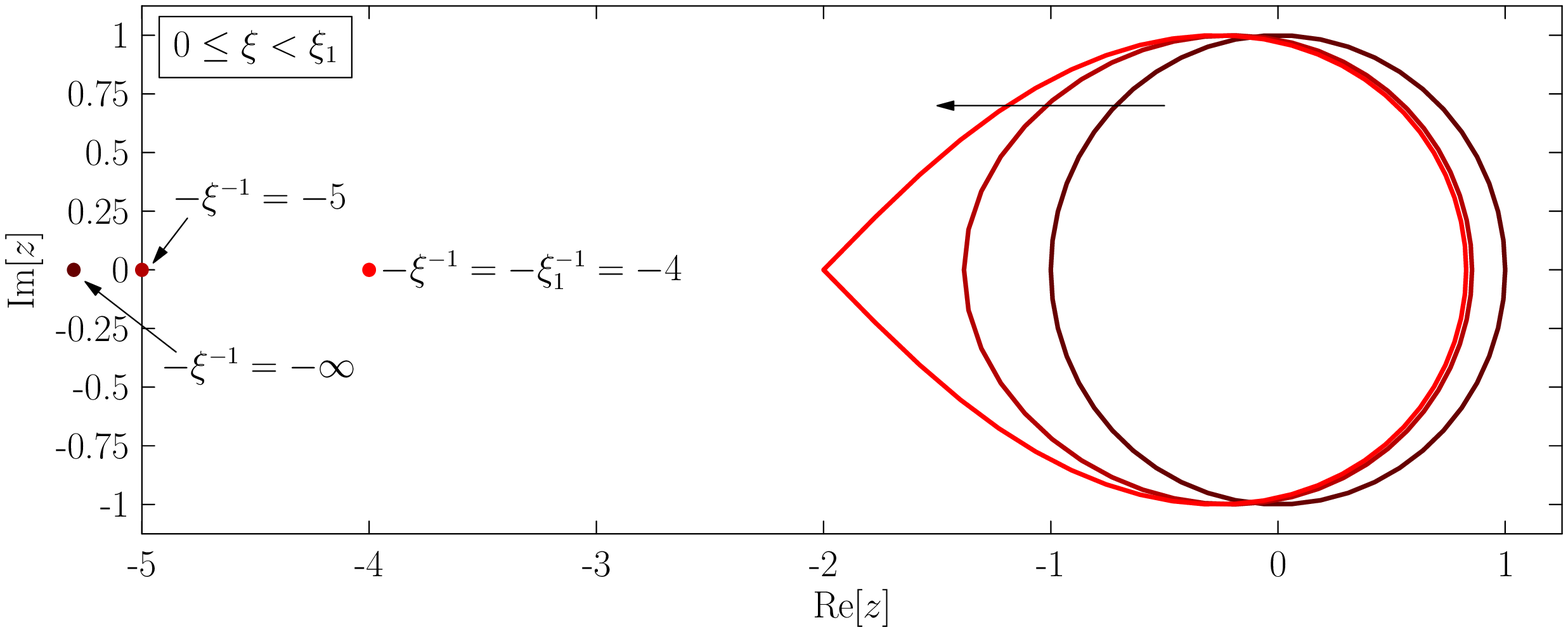}}
\centerline{\includegraphics[width=5.9in]{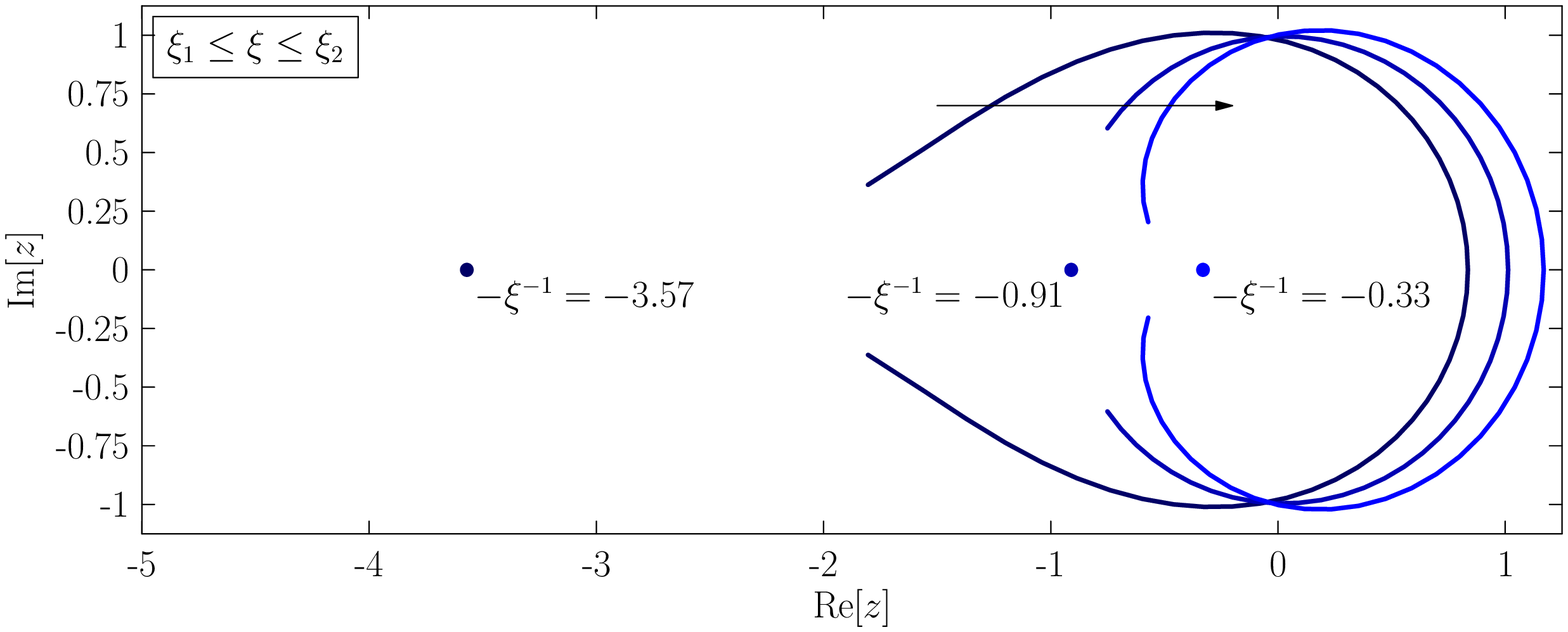}}
\centerline{\includegraphics[width=5.9in]{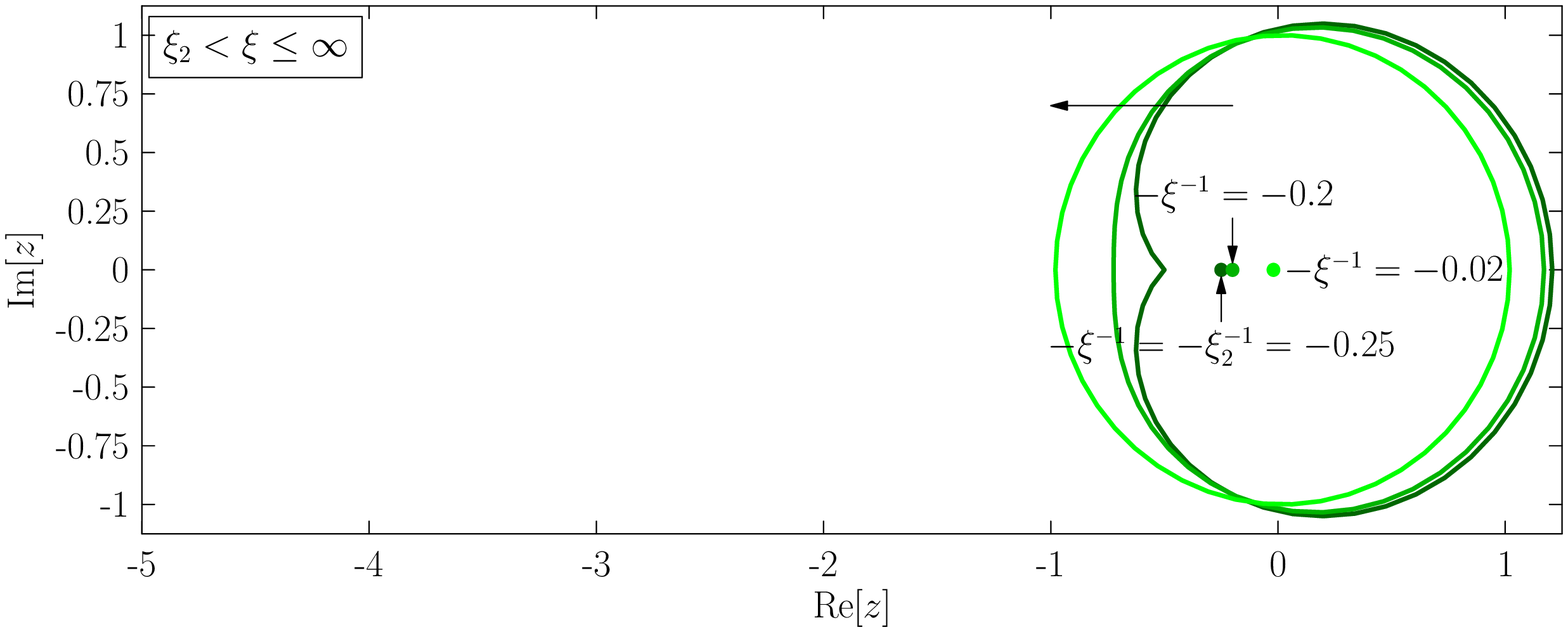}}
\caption{\small The contour ${\cal C}$, which gives the distribution of the eigenvalues of the Polyakov line, showing the transition from the
  small $\xi$ closed phase (in red), the open phase (in blue)
  and the large $\xi$ closed phase (green). }\label{fig1}
\end{figure}

\begin{figure}[ht]
\centerline{\includegraphics[width=10cm]{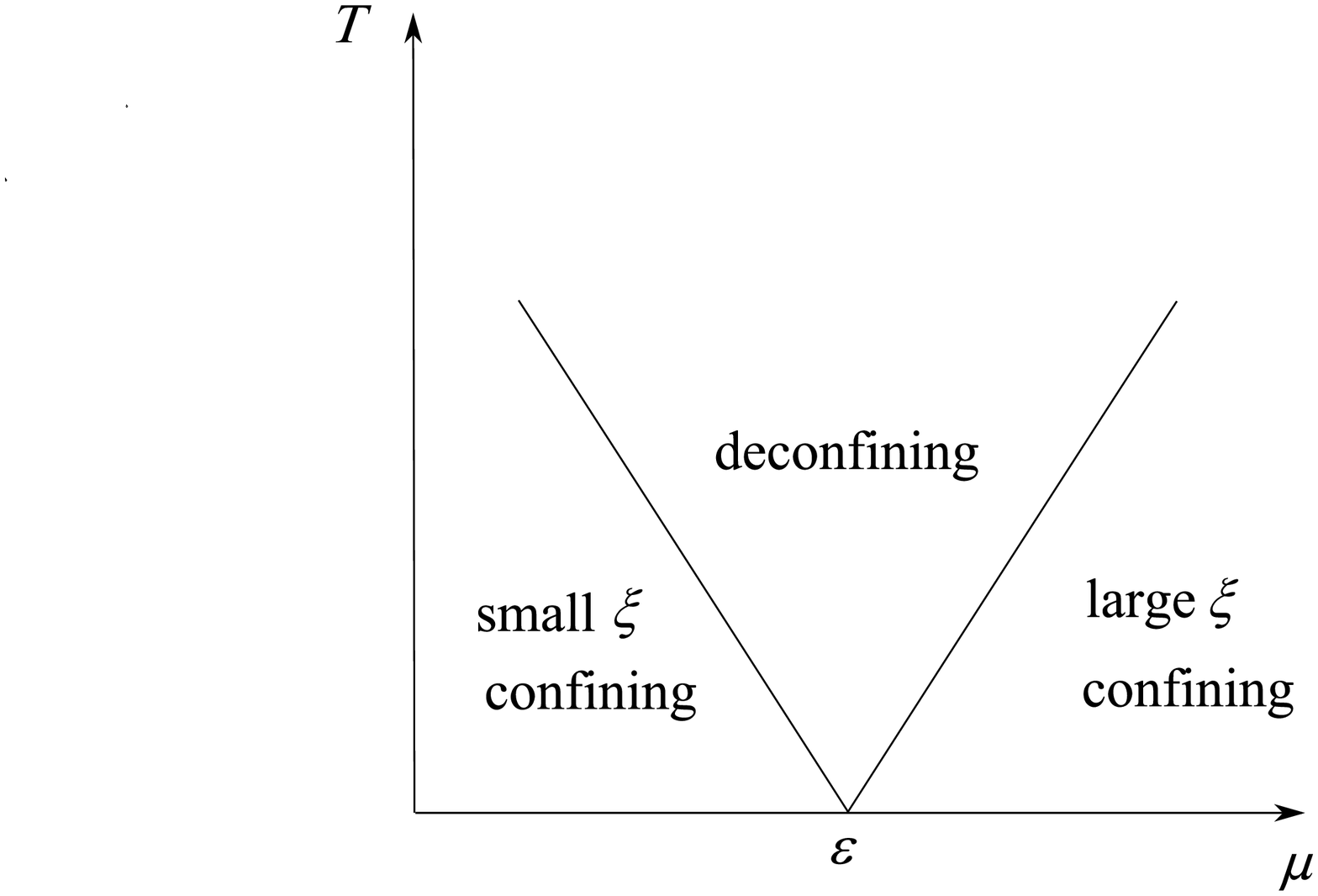}}
\caption{Phase structure in the single level model. For finite $T$ as
  $\mu$ is increased there is a Gross-Witten transition to the
  deconfined phase and then another Gross-Witten transition back to
  the confining phase at a larger $\mu$. In the process the pole at
  $z=-\frac1\xi$ goes from outside the contour ${\cal C}$ to inside.}
\label{f2}
\end{figure}

The behaviour of the contour ${\cal C}$ is shown in the three plots
that make up Figure \ref{fig1}. These plots also show how the pole at
$z=-\tfrac1\xi$ is ``eaten'' as one goes from the small to large $\xi$
phases. Finally Figure \ref{f2} shows schematically the phase diagram
in the $(\mu,T)$ plane in the vicinity of $\mu=\varepsilon$ and at low
temperature.

\subsection{Multi-level model}

In the full QCD model there are, of course a series of energy levels as in
\eqref{act3}. For a series of levels we can write
\EQ{
V(\theta)=i\kN\theta-\sum_\ell\sigma_\ell\log\left(1+\xi_\ell
  e^{i\theta}\right)\ ,
}
with 
\EQ{
\xi_\ell=e^{\beta(\mu-\varepsilon_\ell)}\ .
}

In general a matrix model with such a potential could exhibit a rich
set of phases where the eigenvalues have support over a multiple set of
open contours, the ``multi-cut'' solutions. These multi-cut solutions
arise when the potential has multiple degenerate minima and
the cuts are located around the minima. In the present
case the potential apparently does not exhibit multiple minima and so
we suspect that these multi-cut solutions never dominate the
ensemble. Henceforth, we will only consider solutions with a
closed contour, the confining phase, and with single cuts, the
deconfined phase, leaving a detailed analysis of multi-cut solutions for
the future. 

When $T$ is low enough, experience
with the single-level model suggests that as $\mu$ increases the system
goes through a series of phase transitions with a confined phase
lying in a window of $\mu$ with
$\varepsilon_\ell<\mu<\varepsilon_{\ell+1}$,  for which the closed contour
${\cal C}$ contains the poles $\xi_{\kappa}$, with $\kappa\leq\ell$. In
this case the saddle-point equation is solved by 
\EQ{
\varrho(z)=
\frac{\kN+1}z-\sum_{\kappa\leq\ell}\frac{\xi_{\kappa}
\sigma_{\kappa}}{1+\xi_{\kappa} z}+
\sum_{\kappa>\ell}\frac{\xi_{\kappa}\sigma_{\kappa}}{1+\xi_{\kappa} z}\ .
}
with the effective fermion number 
\EQ{
\kN=\sum_{\kappa\leq\ell}\sigma_{\kappa}
}
indicating the first $\ell$ levels are filled. 

\begin{figure}[t]
\centerline{\includegraphics[width=10cm]{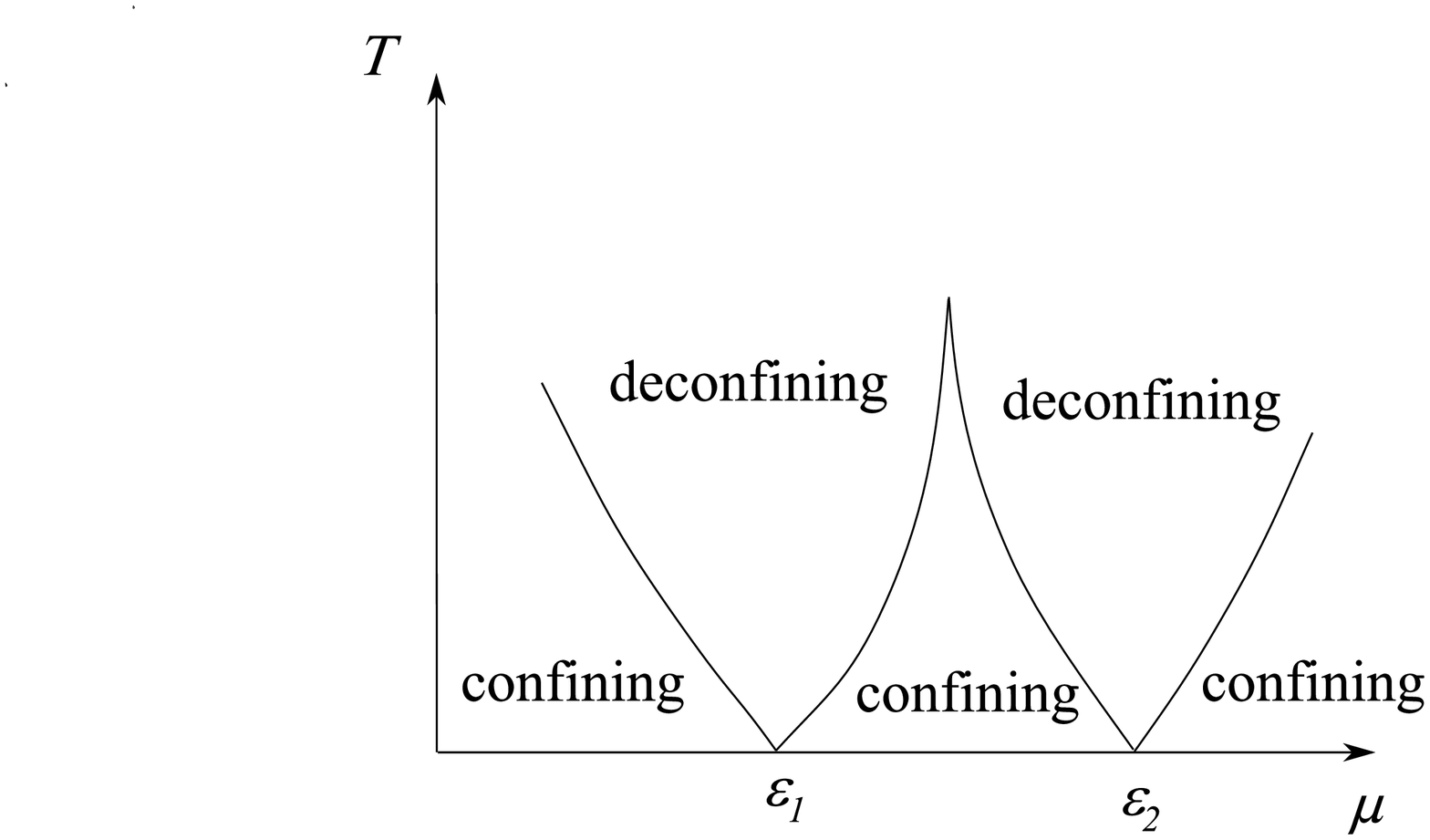}}
\caption{A schematic picture of the phase structure of the multi-level
  model. The 3 portions of confining phase 
are characterized by the fact that the
  closed contour ${\cal C}$ contains the poles at $\{0\}$,
  $\{0,-\frac1{\xi_1}\}$ and $\{0,-\frac1{\xi_1},-\frac1{\xi_2}\}$,
  respectively, as $\mu$ increases.
}
\label{f3}
\end{figure}

The actual distribution of eigenvalues follows from integrating
\eqref{dffe} which gives
\EQ{
e^{is}=z\frac{\prod_{\kappa>
    \ell}(1+\xi_{\kappa}z)^{\sigma_{\kappa}}}{\prod_{{\kappa}\leq{\ell}}
((\xi_{\kappa}z)^{-1}+1)^{\sigma_{\kappa}}}\ .
}
The phase only exists over the finite region of $\mu$ for which
$z(\pi)=z(-\pi)$ so that the contour ${\cal C}$ is actually 
closed. The boundary
of this region can be obtained by the condition that on the boundary
$\varrho(z)=0$ at some point on $z\in{\cal C}$. This defines a triangular
shaped region in the $(\mu,T)$ plane which at $T=0$ 
covers the region
$\varepsilon_{\ell}\leq\mu\leq\varepsilon_{\ell+1}$. 
The apex of
the triangle lies at a special point where $\varrho(z)$ has a double
zero on ${\cal C}$. So with a number of energy levels the boundary of
the confined phases describes a saw-tooth pattern in the $(\mu,T)$
plane as illustrated in Figure \ref{f3}.

\subsection{The continuum model}

In the limit, $m\gg R^{-1}$, the fermionic levels form an approximate
continuum starting at $\varepsilon=m$ and extending up to roughly $2m$
where the discrete structure starts to manifest itself again. As long
as the temperature is not too low, $T\gg\frac1{mR^2}$ (along with
$T\ll\frac1R$ required to have a unitarity matrix model) the sum over
levels can be approximated by a continuum of the form
\EQ{
V(\theta)=i\kN\theta-\frac{2\sigma}{\sqrt\pi}\int_0^\infty dy\,\sqrt y
\log\big(1+\xi e^{-y}e^{i\theta}\big)\ ,
}
where $\xi=e^{\beta(\mu-m)}$ and
$\sigma=\sqrt{2 \pi} \frac{N_f}{N}(R^2mT)^{\frac32}$, with $\sigma\gg1$. In this case, 
\EQ{
zV'(z)=\kN-\frac{2}{\sqrt{\pi}}\int_0^\infty dy\,\frac{\sqrt y\,\sigma\xi e^{-y}z}{1+\xi
  e^{-y} z}=\kN+\sigma\text{Li}_{\frac32}(-\xi z)\ .
\label{zvp}
}

Based on what we have learnt hitherto, we expect
that in this case there is a confined phase for small
enough $\mu$ and then a transition to a deconfined phase at
$\mu\simeq m$. In this case for
larger $\mu$ the deconfined phase persists since the cut in 
\eqref{zvp} does not allow the end-points of the open distribution
to close up again. So the validity of the continuum approximation
implies a $T$ which is large enough so that there is no series of transitions.
The confined phase is simple to analyse using the
techniques in earlier sections. One finds that $\kN=0$ and the density is
\EQ{
\varrho(z)=\frac1z-\frac\sigma z\text{Li}_{\frac32}(-\xi z)\ .
}
The contour ${\cal C}$ is determined by inverting
\EQ{
e^{is}=ze^{-\sigma\text{Li}_{\frac52}(-\xi z)}\ .
}
The phase persists up to a value of $\mu$ such that the zero of
$\varrho(z)$ at, say $z=z_0$ ($\IM z_0=0$ and $z_0<0$), 
lies on the contour ${\cal C}$, which
means that $z_0$ and $\xi$ are roots of 
\EQ{
1-\sigma\text{Li}_{\frac32}(-\xi z_0)=0~~~~~~
\text{and}~~~~~~
z_0e^{-\sigma\text{Li}_{\frac52}(-\xi z_0)}=-1\ .
\label{koo}
} 
Since $\sigma\gg1$, the solution is approximately $z_0=-e$ and $\xi
=(e\sigma)^{-1}$ and so the transition occurs at
\EQ{
\mu=m-T\log(e\sigma)=m-T\Big[1+\frac32\log(R^2mT)+\frac12\log
2 \pi + \log\Big(\frac{N_f}{N}\Big)\Big]\ .
\label{hss}
}

The deconfined phase in this case is described by a resolvent
\EQ{
\omega(z)=\kN-\frac{2\sigma}{\sqrt\pi}\int_0^\infty dy\,
\frac{\sqrt y\,\xi e^{-y}}{1+\xi
  e^{-y} z}\left[z+
\frac{\sqrt{(z-\tilde z)(z-\tilde z^*)}}{\big|1+\xi e^{-y}\tilde z\big|}
\right]\ .
}
In principle, one can find $\tilde z$ numerically,
however, we don't pursue this analysis here.

\section{The Large $N$ Theory in the $(\mu,T)$ Plane}

When we move out of the regime of low temperature, so that $T$ is no
longer $\ll R^{-1}$, into the whole $(\mu,T)$ plane, the factors 
$z_b(n\beta/R)$ in
\eqref{pop} cannot be ignored. Hence, the measure on $P$ can
no longer be approximated by that of a unitary matrix.   
In this case, we can in principle solve the model in the confining
phase, where the eigenvalues lie on a closed contour by introducing
the Laurent expansion
\EQ{
\rho(z)=\sum_{n=-\infty}^{\infty} \rho_n z^{-n-1}\ ,
}
where $\rho_0=1$ follows from the normalization condition. 
In terms of the components, the action of
the Polyakov line with $N_f$ fundamental fermions is
\SP{
S(\rho_n)&=N^2\sum_{n=1}^\infty\frac1n
\Big[(1-z_b(n\beta/R))\rho_n\rho_{-n}\\
&~~~~~~~+\frac{N_f}{N}(-1)^n z_f(n\beta/R,mR)\Big(e^{n\beta\mu}\rho_n+
e^{-n\beta\mu}\rho_{-n}\Big)\Big]\ .
\label{act5}
}
The action is quadratic in the $\rho_n$ and so the saddle-point is
easily found to be at
\EQ{
\rho_n=\frac{N_f}N(-1)^{n+1}e^{-n\beta\mu}\frac{z_f(\left| n\right|\beta/R,mR)}{1-z_b(\left| n\right|\beta/R)}\ ,
} 
for $n\neq0$ and $\rho_0=1$. 
Note that the $SU(N)$ condition \eqref{con} is automatically
satisfied.
The boundary of the confining phase is then obtained by 
solving for the condition that $\rho(z_0)=0$ for $z_0$ lying on the
contour ${\cal C}$ which itself is determined from \eqref{dffe}.

When $m\gg R^{-1}$, and in the intermediate regime
\EQ{
\frac1R\gg T\gg \frac1{m R^2}
\label{reg1}
}
we have already determined that there is a Gross-Witten 
transition 
\eqref{hss}. At $\mu=0$ we know that, for $m\gg\frac1R$, the fermionic
modes are decoupled. So we have effectively the pure gauge theory which
is known to have a
confinement/deconfinement transition driven by the modes
$\rho_{\pm1}$ at a temperature where $z_b(\beta/R)=1$, giving
$T\simeq0.759R^{-1}$ \cite{Aharony:2003sx}. This transition is known
to be first order even when higher order corrections in the coupling
are considered; in fact the first correction to the
Gaussian approximation requires a 3-loop computation
\cite{Aharony:2005bq}. For finite fermion mass the transition occurs
below $0.759R^{-1}$. If we make the approximation of 
ignoring the higher modes $\rho_n$,
$|n|>1$, then the transition happens when $\rho_1 = \rho_{-1} = 1/2$ such that
\EQ{
\frac{1}{2}z_b(\beta/R)+\frac{N_f}{N}z_f(\beta/R,mR)=\frac{1}{2}\ ,
}
and is a third order Gross-Witten transition. The case with $m=0$ was
considered in \cite{Schnitzer:2004qt}.

\begin{figure}[t]
\centerline{\includegraphics[width=10cm]{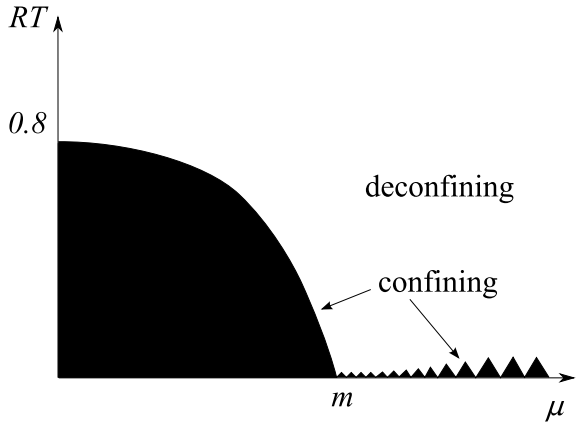}}
\caption{A schematic picture of the phase structure of theory with
  $m\gg\frac1R$. The discrete structure of the energy levels is only
 apparent when $T\ll\frac1{m R^2}$ which we have indicated by the
  sawtooth pattern.}
\label{f4}
\end{figure}

The Gross-Witten transition at $\mu=0$ extends out
into the $(\mu,T)$ plane and joins smoothly with the transition line that
comes from the first energy level.
Although the line of transitions can be
determined numerically a good approximation is obtained by keeping
only the first 2 modes $\rho_{\pm1}$. A schematic picture of the phase
diagram for the large mass regime is shown in Figure \ref{f4}. We have
also indicated on this diagram the discrete sawtooth structure which can be
seen at very low temperatures $T\ll\frac1{m R^2}$. In this regime the
peaks rise until $\mu$ approaches $\sim2m$ and then they decrease slowly.

\begin{figure}[t]
\centerline{\includegraphics[width=10cm]{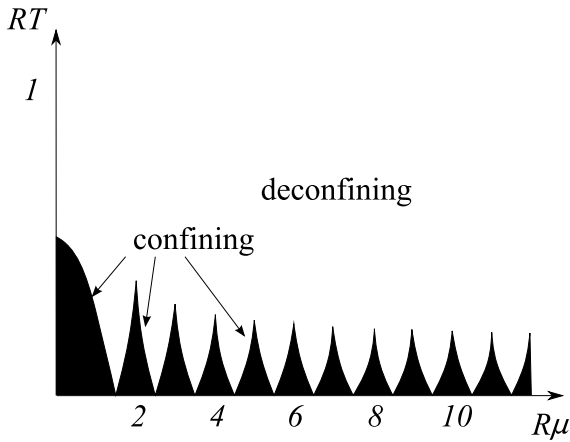}}
\caption{A schematic picture of the phase structure of theory with
  $m=0$ based on numerical approximations. 
The deconfined phases touch the $T=0$ axis at the points
  $\mu=(\ell+\tfrac12)R^{-1}$, $\ell=1,2\ldots$.}
\label{f5}
\end{figure}

When $m$ is reduced, the transition at $\mu=0$ occurs at a lower
temperature and a schematic picture of the phase diagram
appears in Figure \ref{f5}. Now the sawtooth structure extends all the
way to small $\mu$.


\section{Comparison of $N = 3$ and $\infty$}

\begin{figure}[t]
  \hfill
  \begin{minipage}[t]{.49\textwidth}
    \begin{center}
\includegraphics[width=0.99\textwidth]{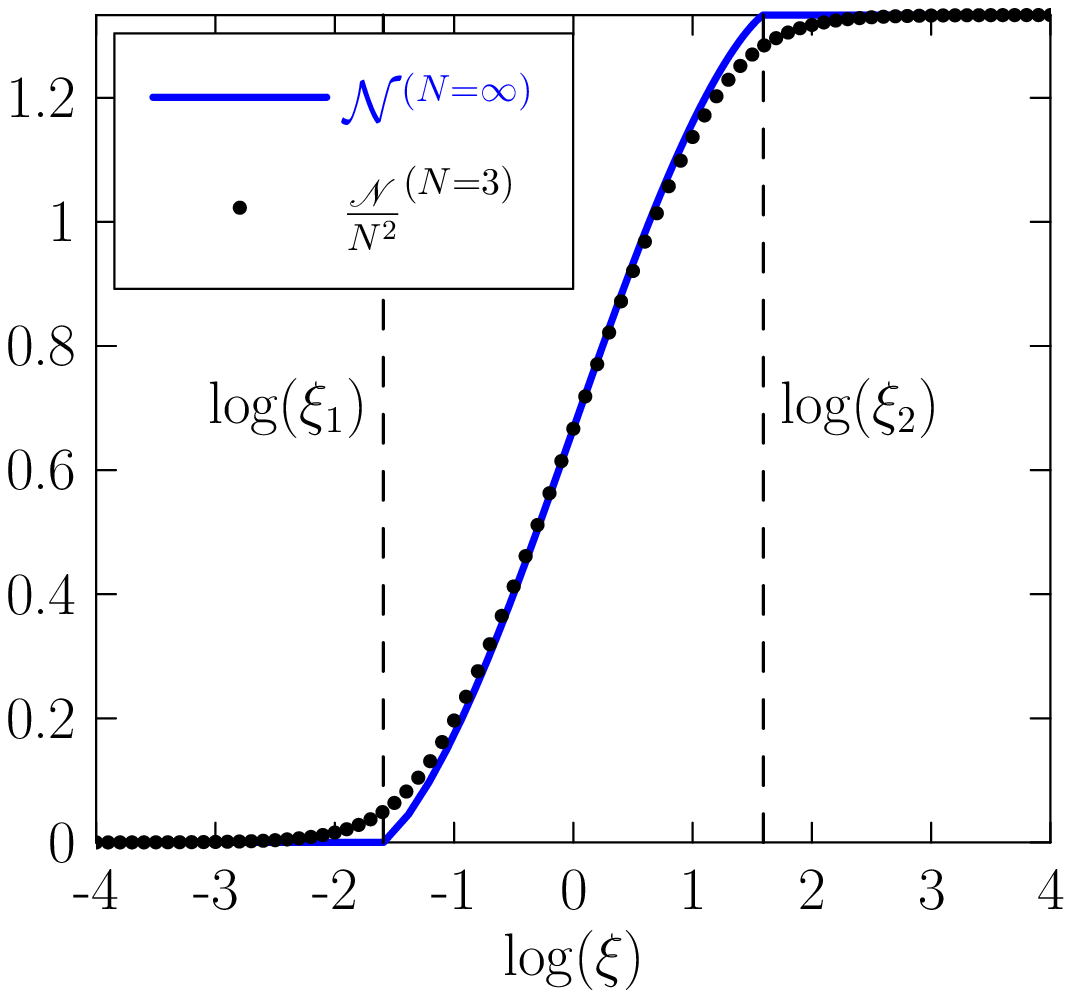}
    \end{center}
  \end{minipage}
  \hfill
  \begin{minipage}[t]{.49\textwidth}
    \begin{center}
\includegraphics[width=0.99\textwidth]{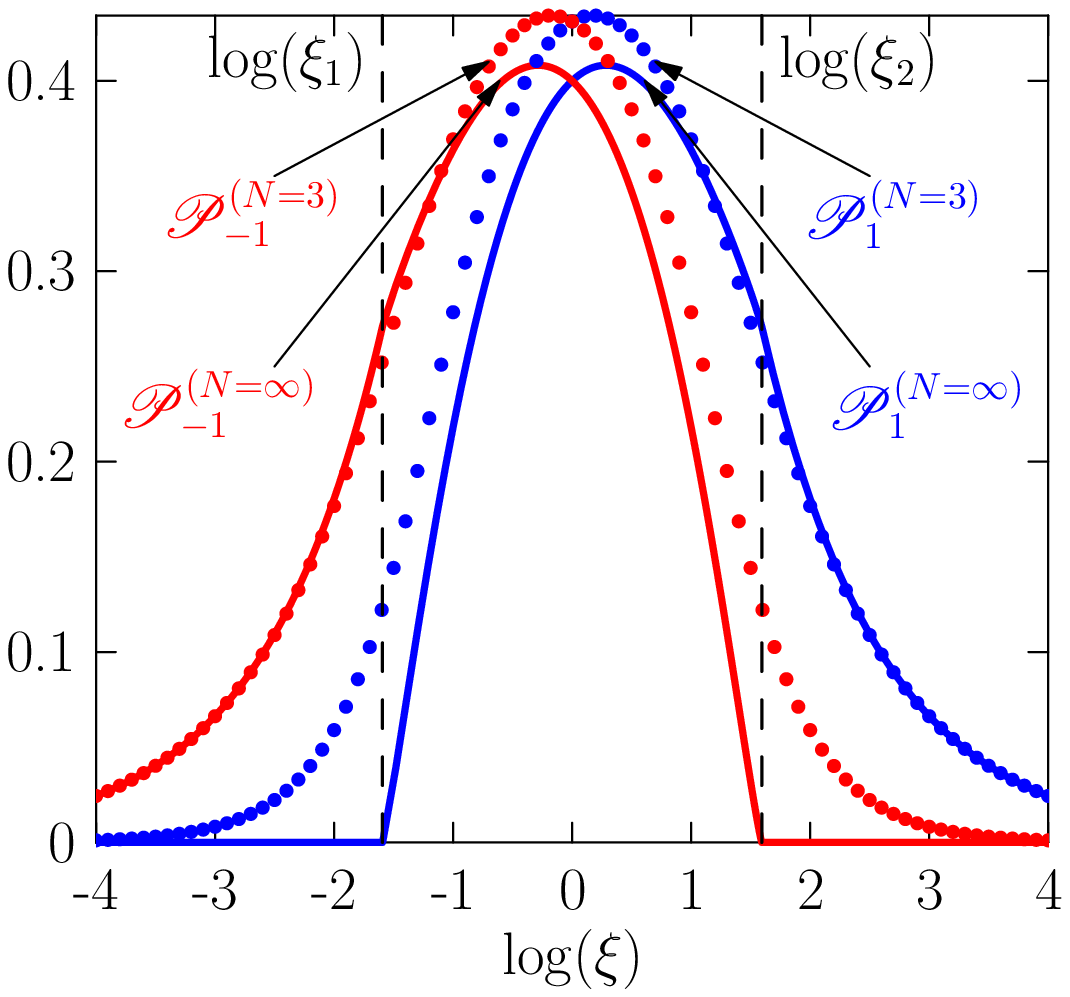}
    \end{center}
  \end{minipage}
  \hfill
\caption{Comparison of $N = 3$ and $N = \infty$ QCD at the first level transition ($l = 1$) with $m R = 0$, $\sigma_1 = \frac{4}{3}$. (Left): Fermion number. (Right): Polyakov lines $\PP_{1}$ and $\PP_{-1}$.}
\label{comp_n3_inf}
\end{figure}

It is meaningful to compare the results we have obtained using the two
different calculational techniques presented in this paper. This is
useful not only towards checking our results, but also towards
learning about what differences might arise when considering small
vs. large $N$, and under which conditions there is agreement. 

We consider a single level transition, $\ell = 1$, and $\sigma =
\frac{4}{3}$, which is the value used in the $N = 3$ calculations. In
Figure \ref{comp_n3_inf} (Left) we show an overlay of the $N = 3$ and
$\infty$ results for the fermion number normalized by $N^2$. The
transitions between the small $\xi$ confined phase and the deconfined
phase, and also between the deconfined phase and the large $\xi$
confined phase, are clearly signified by discontinuities for $N =
\infty$, but they are smoothed out for $N = 3$. Taking lower
temperatures does not sharpen the $N = 3$ result since we are plotting as a function
of $\log \xi$ rather than $\mu$. It may be that the transitions are smoother
for $N = 3$ as a result of working in a small spatial volume. The important feature of
this comparison is that the $N = 3$ calculation suggests possible
transitions and the $N = \infty$ calculation confirms them. 

In Figure \ref{comp_n3_inf} (Right) we show the Polyakov lines $\PP_1$
and $\PP_{-1}$ for $N = 3$ and $\infty$. Again the discontinuities
which occur at the transitions in the $N = \infty$ case are lost for
$N = 3$. In addition, since the difference $\left| \PP_1 - \PP_{-1}^*
\right|^{(N=\infty)} \gapprox \left| \PP_1 - \PP_{-1}^*
\right|^{(N=3)}$, and given our observation that the severity of the
sign problem increases with $\left| \PP_1 - \PP_{-1}^* \right|$, then it
appears that the sign problem would always be more severe in the large $N$
limit. 

\begin{figure}[t]
  \hfill
  \begin{minipage}[t]{.32\textwidth}
    \begin{center}
\includegraphics[width=0.99\textwidth]{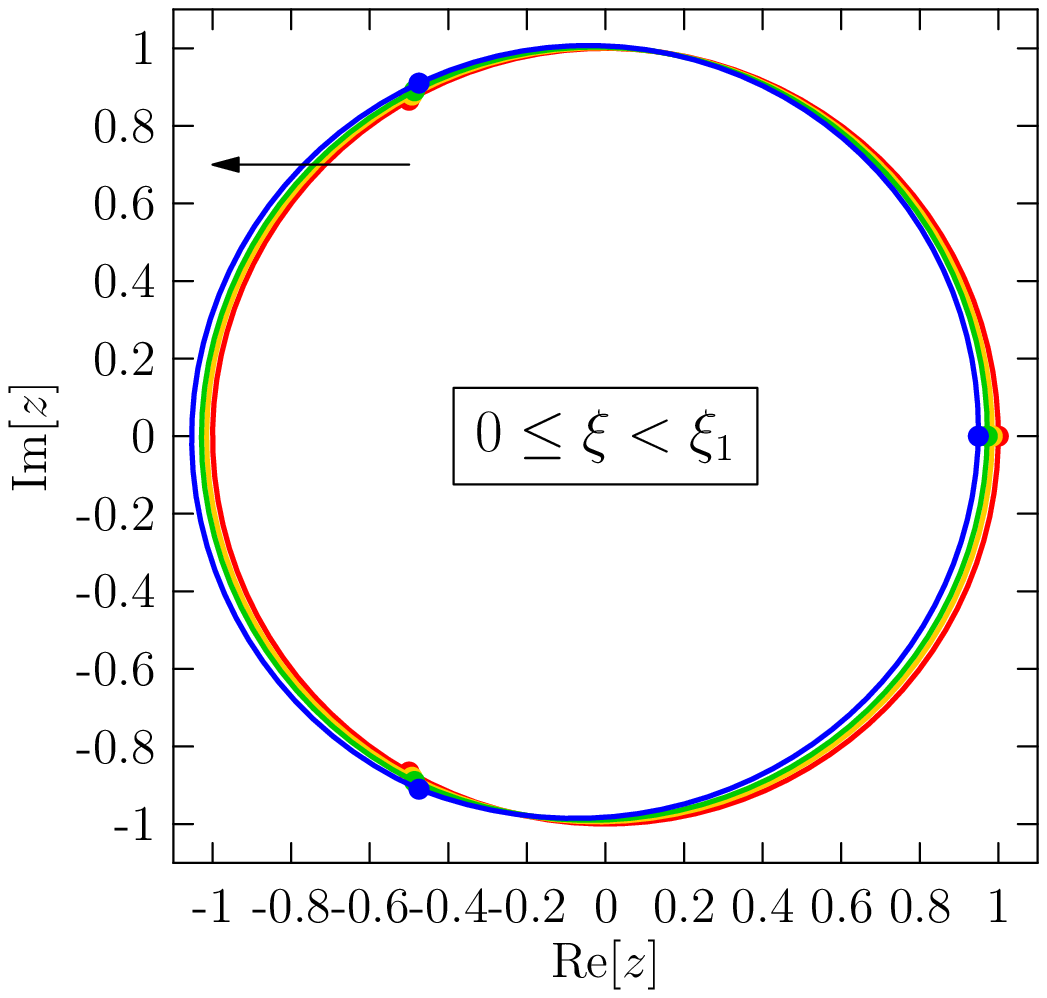}
    \end{center}
  \end{minipage}
  \hfill
  \begin{minipage}[t]{.32\textwidth}
    \begin{center}
\includegraphics[width=0.99\textwidth]{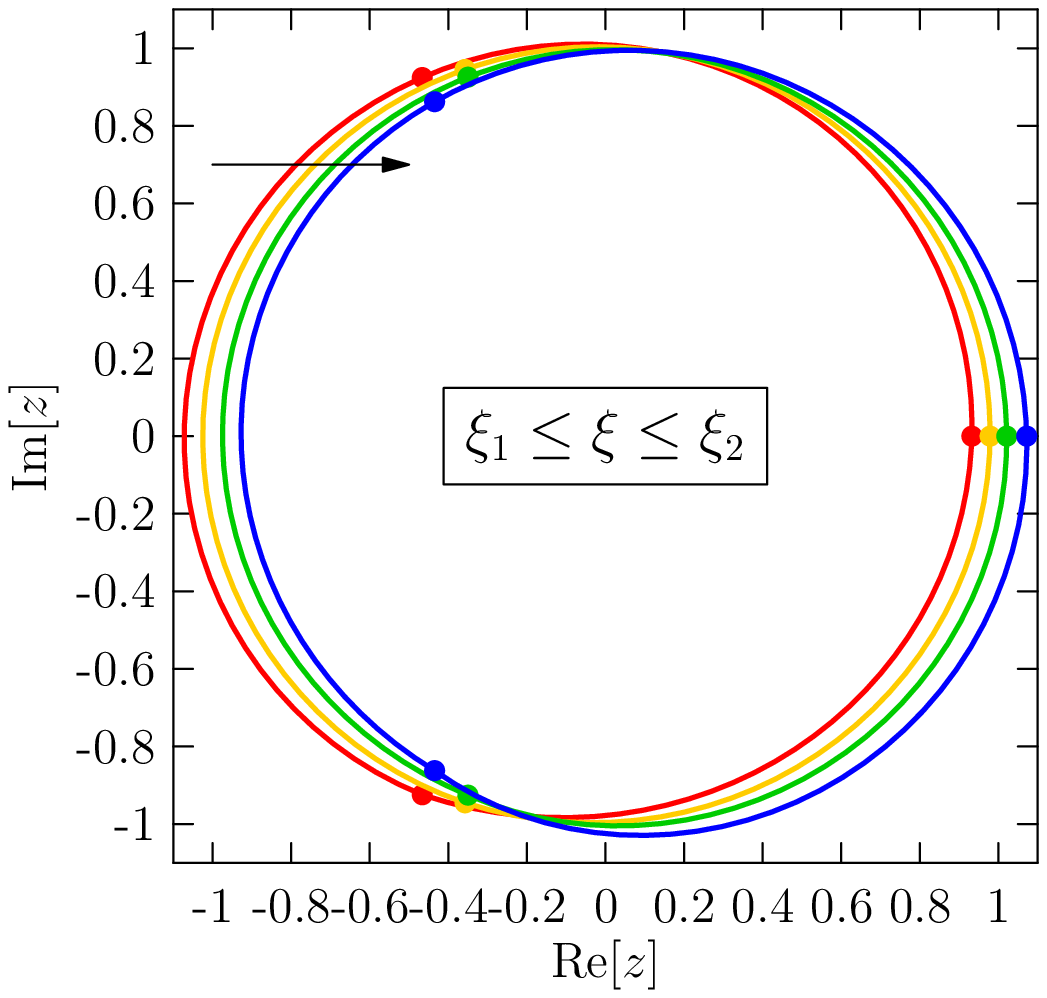}
    \end{center}
  \end{minipage}
  \hfill
  \begin{minipage}[t]{.32\textwidth}
    \begin{center}
\includegraphics[width=0.99\textwidth]{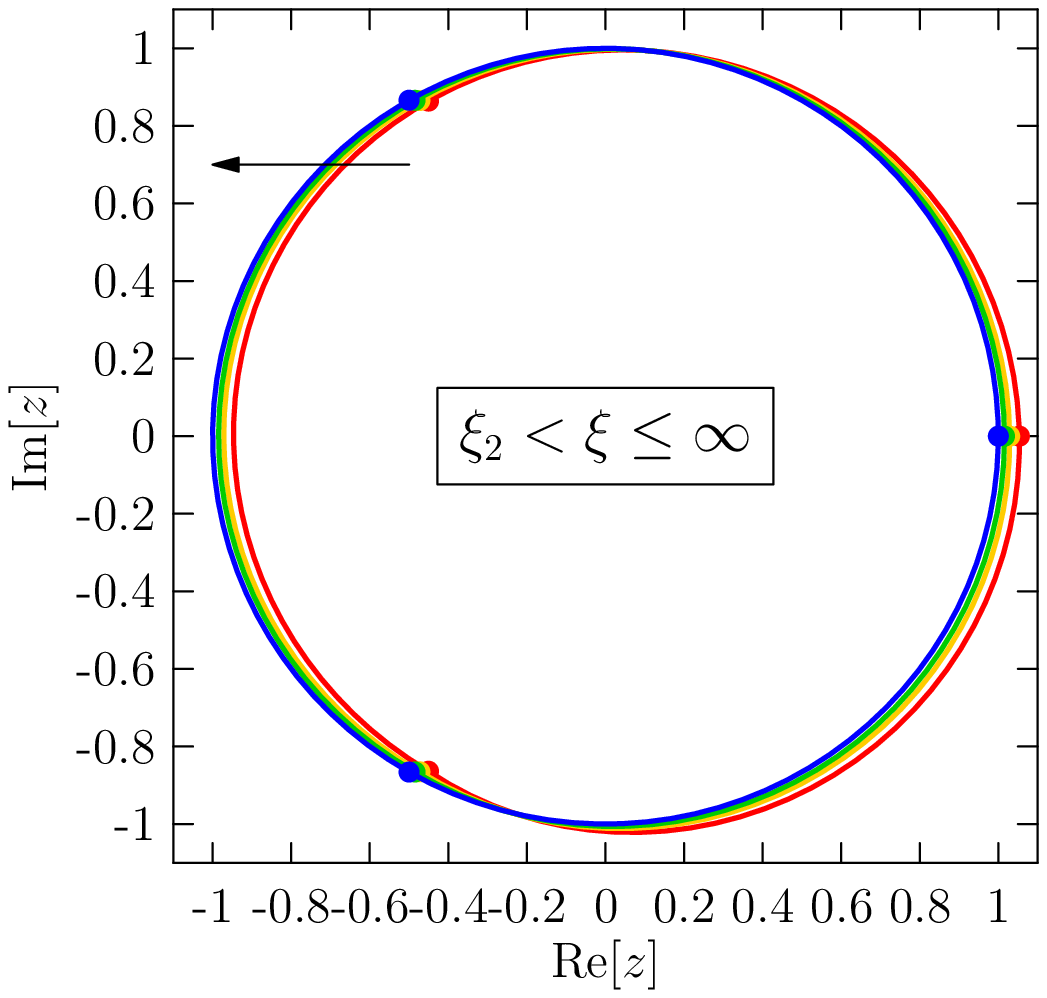}
    \end{center}
  \end{minipage}
  \hfill
\caption{The points in each figure represent the distribution values
  $z_i = e^{i \phi_i}$ of the angle expectation values $\phi_i$ of the
  Polyakov line for $N = 3$. Four values of $\xi$ in each of the
  regions $\xi < \xi_1$, $\xi_1 \le \xi \le \xi_2$, and $\xi > \xi_2$
  are considered. The contours connecting the points are
  interpolations between them and show how the distribution moves with
  $\xi$ (in each plot as $\xi$ increases the contours change from
  red$\rightarrow$yellow$\rightarrow$green$\rightarrow$blue). This is
  in qualitative agreement with the large $N$ distributions.}
\label{zdist_n3}
\end{figure}

It is also possible to solve for the distribution $z_i = e^{i \phi_i}$
from the expectation values $\phi_i$ of the angles of the Polyakov
line for $N = 3$, for the purpose of comparing with $N = \infty$. To
do this we solve the characteristic equation 
\EQ{
(e^{i \theta_1} - z)(e^{i \theta_2} - z)(e^{-i (\theta_1 +\theta_2)} - z) = 0 .
}
This result can be expanded and rewritten in terms of the expectation
values $\PP_1 = e^{i \theta_1} + e^{i \theta_2} + e^{-i (\theta_1 +
  \theta_2)}$, and $\PP_{-1} = e^{-i \theta_1} + e^{-i \theta_2} +
e^{i (\theta_1 + \theta_2)}$, determined earlier, to give 
\EQ{
z^3 - z^2 \PP_{1}+ z \PP_{-1} - 1 = 0 .
}
We find the three solutions for $z$ (using our results for $\PP_1$ and
$\PP_{-1}$ at the $l = 1$ level transition) for four different values
of $\xi$ in each of the regions $\xi < \xi_1$, $\xi_1 \le \xi \le
\xi_2$, and $\xi > \xi_2$. The results are given by the points in
Figure \ref{zdist_n3}. The contours are interpolations of the points
and give a rough idea of how the distribution changes with
$\xi$. Comparing with the large $N$ results for the distribution,
given in Figure \ref{fig1}, we see that the direction of movement of
the distribution matches up in all three regions: In the complex
$z$-plane, for $\xi < \xi_1$ the contour passes from the unit circle
to the left with increasing $\xi$, for $\xi_1 \le \xi \le \xi_2$ it passes
back to the right passing by the unit circle, and for $\xi > \xi_2$
the distribution moves left again back to the unit circle. 

It is also worth noting that there is agreement between the large $N$
finite temperature phase diagrams in Figures \ref{f4} and \ref{f5},
and the low temperature results for the Polyakov line for $N = 3$ in
Figures \ref{polys}, and \ref{poly_phase_trans_mass} (Left). In the
case of a massless quark, and at low temperatures, we see a widening
of the deconfined phases for increasing $\mu$. For a massive quark,
at low but non-zero temperature, there is a brief respite from the
oscillating phases after the onset transition to the first deconfined phase
at $\mu = m$, but taking $\mu$ large enough causes the oscillations to
return. 


\section{Conclusions and Outlook}

We have performed a one-loop analytical derivation of the phase
diagram of QCD as a function of temperature and chemical potential, in
the large $N$ and $N_f$ limit, on $S^1 \times S^3$, and supported the
low temperature results with numerical calculations for $N = 3$. In
the case of massless quarks, and considering the low temperature
limit, we observe a series of confinement-deconfinement transitions as
a function of the chemical potential. In the large $N$ limit the
phases are characterized by the distribution of the Polyakov line
eigenvalues in the complex plane which can be obtained using matrix
model techniques that have been generalized for a complex action. In the large quark mass limit we observe the ``Silver Blaze" feature in that bulk observables are roughly zero until the onset transition to the deconfined phase which occurs at $\mu = m$. From here there is a brief continuum-like behavior in that the observables appear smooth after the transition. This behavior continues until $\mu$ is sufficiently larger than $m$ that the levels spread out and the confinement-deconfinement phase oscillations return.

From a physical standpoint, a remarkable aspect of our results is the
correlation between deconfinement and the existence of partially filled quark
shells in a box. From the behaviour of the fermion number at low
temperatures we observe that each confinement-deconfinement transition
in the Polyakov line is associated with a level transition. All energy
levels below the Fermi energy, given by the chemical potential, are
filled, and levels above the Fermi energy are empty. There are clear
analogies with partially-filled bands in condensed 
matter physics; here we would say that the Fermi energy in such a
system falls within 
a band, resulting in a conducting ground state.
A more accurate analogy, since it involves a finite number of particles, would
be partially-filled shells in nuclear physics. The lesson we draw is that
de-confinement appears to require a non-zero density of gapless states. We also
learn that these states are either particle-like or hole-like depending on
which edge of
the band is closer. Because particles and holes carry conjugate representations
of the gauge group the resulting
physics is distinct, as revealed by the differing behaviours of $\PP_1$ and
$\PP_{-1}$. The non-monotonic behaviour of
$\PP(\mu)$ has been observed in lattice simulations of QCD with gauge group
SU(2) near its saturation density ({\it i.e.\/}~$2N_cN_f$ quarks per lattice
site)~\cite{Hands:2006ve}.

There are a number of interesting generalizations that might be
made. One could consider more quark flavours with different masses
$m_f$, perhaps coupled to different chemical potentials $\mu_f$. A detailed multidimensional phase diagram
could be calculated as a function of the quark masses and their
chemical potentials. There
may be a sign of color superconducting phases from the configurations
of the gauge field. It might be interesting to consider different
manifolds as well. In addition, a numerical computation of the line of
transitions in the $(\mu,T)$ plane can be performed from the results
for the large $N$ calculation to connect the transition points on the
$T$ and $\mu$ axes for large quark mass, and the curvature of this
line can be determined. In the event that lattice simulations of QCD
become possible at moderate chemical potential in the low temperature
limit, it would be interesting to see if the low temperature
confinement-deconfinement phase oscillations we have observed as a
function of $\mu$ are present at strong coupling and/or large volume. We are currently
investigating this possibility in simulations of QCD with $N = 2$. 

Our analysis has been limited to the Gaussian approximation and
it is clearly important to consider systematically the
effect of higher orders in the gauge coupling. This kind of higher
order analysis was
qualitatively undertaken in the thermal ${\cal N}=4$ theory in
\cite{Aharony:2003sx}. It turns out that in this case the fate of the
confinement/deconfinement transition depends critically on the sign
of the coefficient of a 3-loop term in the expansion of the
effective action. Whilst this 3-loop calculation has not been
performed for the ${\cal N}=4$ gauge theory it has for the pure gauge
theory in \cite{Aharony:2005bq}. This {\it tour
  de force\/} calculation proves that the transition survives as a
first order transition.
The effect of higher orders in the coupling for the theory with
fundamental flavours (with $\mu=0)$ have been considered qualitatively in
\cite{Schnitzer:2006xz,Basu:2008uc}. This latter work suggests in
the theory with $\mu=0$ that the confinement/deconfinement 
transition disappears and becomes a cross-over as $\frac{N_f}N$ is increased
beyond a certain critical value. It will be interesting to see how
this is altered in the presence of a chemical potential.
It will be interesting to see whether the phase structure that we find
can be related to that of a theory with an AdS/CFT-type gravity
dual. For instance, will the infinite sequence of Gross-Witten
transitions that we see be seen in the dual gravitational description?

\section{Acknowledgements}

We would like to thank Gert Aarts, Prem Kumar, and Rob Pisarski for useful discussions.

\clearpage

\end{document}